\pdfoutput=1

\documentclass[11pt,twoside,a4paper,cmspaper,final,collab]{cms-tdr}
\def\svnVersion{465272P}\def\cmsCernNoTag{CERN-EP-2018-058}\def\cmsCernDate{\today}\def\cmsMessage{Published in the Journal of Instrumentation as \href{http://dx.doi.org/10.1088/1748-0221/13/06/P06015}{\doi{10.1088/1748-0221/13/06/P06015.}}}

\begin{document}\cmsNoteHeader{MUO-16-001}

\hyphenation{had-ron-i-za-tion}
\hyphenation{cal-or-i-me-ter}
\hyphenation{de-vices}
\RCS$HeadURL: svn+ssh://svn.cern.ch/reps/tdr2/papers/MUO-16-001/trunk/MUO-16-001.tex $
\RCS$Id: MUO-16-001.tex 464547 2018-06-14 13:54:53Z rakness $

\providecommand{\NA}{\ensuremath{\text{---}}}

\newlength\cmsTabSkip\setlength{\cmsTabSkip}{1ex}
\newlength\cmsFigWidth
\ifthenelse{\boolean{cms@external}}{\setlength\cmsFigWidth{0.85\columnwidth}}{\setlength\cmsFigWidth{0.4\textwidth}}

\ifthenelse{\boolean{cms@external}}{\providecommand{\cmsLeft}{top\xspace}}{\providecommand{\cmsLeft}{left\xspace}}
\ifthenelse{\boolean{cms@external}}{\providecommand{\cmsRight}{bottom\xspace}}{\providecommand{\cmsRight}{right\xspace}}

\ifthenelse{\boolean{cms@external}}{\providecommand{\CL}{C.L.\xspace}}{\providecommand{\CL}{CL\xspace}}

\newcommand{\Zmm}{\ensuremath{\text{\cPZ} \to \MM}}
\newcommand{\chisqdof}{\ensuremath{\chi^2/\mathrm{dof}}}

\cmsNoteHeader{MUO-16-001}

\title{Performance of the CMS muon detector and muon reconstruction with proton-proton collisions at $\sqrt{s}=13\TeV$}

\date{\today}

\abstract{
The CMS muon detector system, muon reconstruction software, and high-level trigger
underwent significant changes in 2013--2014 in preparation for running at higher LHC
collision energy and instantaneous luminosity.
The performance of the modified system is studied using proton-proton collision data at
center-of-mass energy $\sqrt{s}=13\TeV$, collected at the LHC in 2015 and 2016.
The measured performance parameters, including spatial resolution, efficiency, and
timing, are found to meet all design specifications and are well reproduced by
simulation.
Despite the more challenging running conditions, the modified muon system is found to
perform as well as, and in many aspects better than, previously.
We dedicate this paper to the memory of Prof.\ Alberto Benvenuti, whose work was
fundamental for the CMS muon detector.
}

\hypersetup{
pdfauthor={CMS Collaboration},
pdftitle={Performance of the CMS muon detector and muon reconstruction with proton-proton collisions at sqrt(s) = 13 TeV},
pdfsubject={CMS},
pdfkeywords={CMS, LHC, Muon, Detector, Drift Tubes, RPC, CSC}}

\maketitle

\section{Introduction}
\label{sec:intro}

The Compact Muon Solenoid (CMS) detector at the CERN LHC is a general purpose device
designed primarily to search for signatures of new physics in proton-proton
(\Pp\Pp) and heavy ion (proton-ion and ion-ion) collisions.
Since many of these signatures include muons, CMS is constructed with subdetectors
to identify muons, trigger the CMS readout upon their detection, and measure their
momentum and charge over a broad range of kinematic parameters.
In this paper, the composite whole of muon subdetectors is called the muon detector,
and the software algorithms used to combine the data from all CMS subdetectors to
characterize the physics objects created in collisions are collectively referred to
as particle reconstruction.
Previous published studies of the performance of the CMS muon
detector~\cite{DPGPerformance} and muon reconstruction~\cite{POGPerformance} were
based on data from {\Pp\Pp} collisions at center-of-mass energy $\sqrt{s}=7\TeV$.
These data were collected in 2010, the first full year of LHC operations (the first
year of ``Run~1'',  which lasted from 2010 to 2012).
To prepare for the higher collision energy and luminosity of the subsequent running
period (``Run~2'', beginning in 2015), significant improvements were made to the
muon system in 2013--2014 during the long shutdown period between Runs 1 and~2.
These improvements will be described in Section~\ref{sec:detectors}.
The present paper describes the performance of the Run~2 CMS muon system, and covers
the subdetectors, the reconstruction software, and the high-level trigger.
It is based on data collected in 2015 and 2016 from {\Pp\Pp} collisions at
$\sqrt{s}=13\TeV$ with instantaneous luminosities up to~$8\ten{33}$\percms.
As a result of these improvements to the muon detector and reconstruction algorithms,
and in spite of the higher instantaneous luminosity, the performance of the muon
detector and reconstruction is as good as or better than in 2010.
Moreover, all performance parameters remain well within the design specifications of
the CMS muon detector~\cite{CMSMuonTDR}.

An extensive description of the performance of the muon detector and the muon
reconstruction software has been given in Ref.~\cite{DPGPerformance} and
Ref.~\cite{POGPerformance}.
Therefore, in this paper, representative performance plots from individual muon
subsystems are shown and results from the other subsystems, when pertinent, are
described in the text.
A description of the different subdetectors forming the CMS muon detector is
given in Section~\ref{sec:detectors}.
The muon reconstruction, identification, and isolation algorithms are outlined in
Section~\ref{sec:muonreco}, followed by a short description of the data and
simulation samples used in Section~\ref{sec:samples}.
The performance of individual muon subdetectors and that of the full system is
described in detail, particularly with regard to spatial resolution
(Section~\ref{sec:spatialresolution}), efficiency (Section~\ref{sec:efficiency}),
momentum scale and resolution (Section~\ref{sec:momentumscale}), and timing
(Section~\ref{sec:timing}).
The design and performance of the high-level trigger is described in
Section~\ref{sec:trigger}.
The results are summarized in Section~\ref{sec:summary}.

\section{Muon detectors}
\label{sec:detectors}

\begin{figure}[ht]
  {
    \centering
    \includegraphics[width=0.8\textwidth]{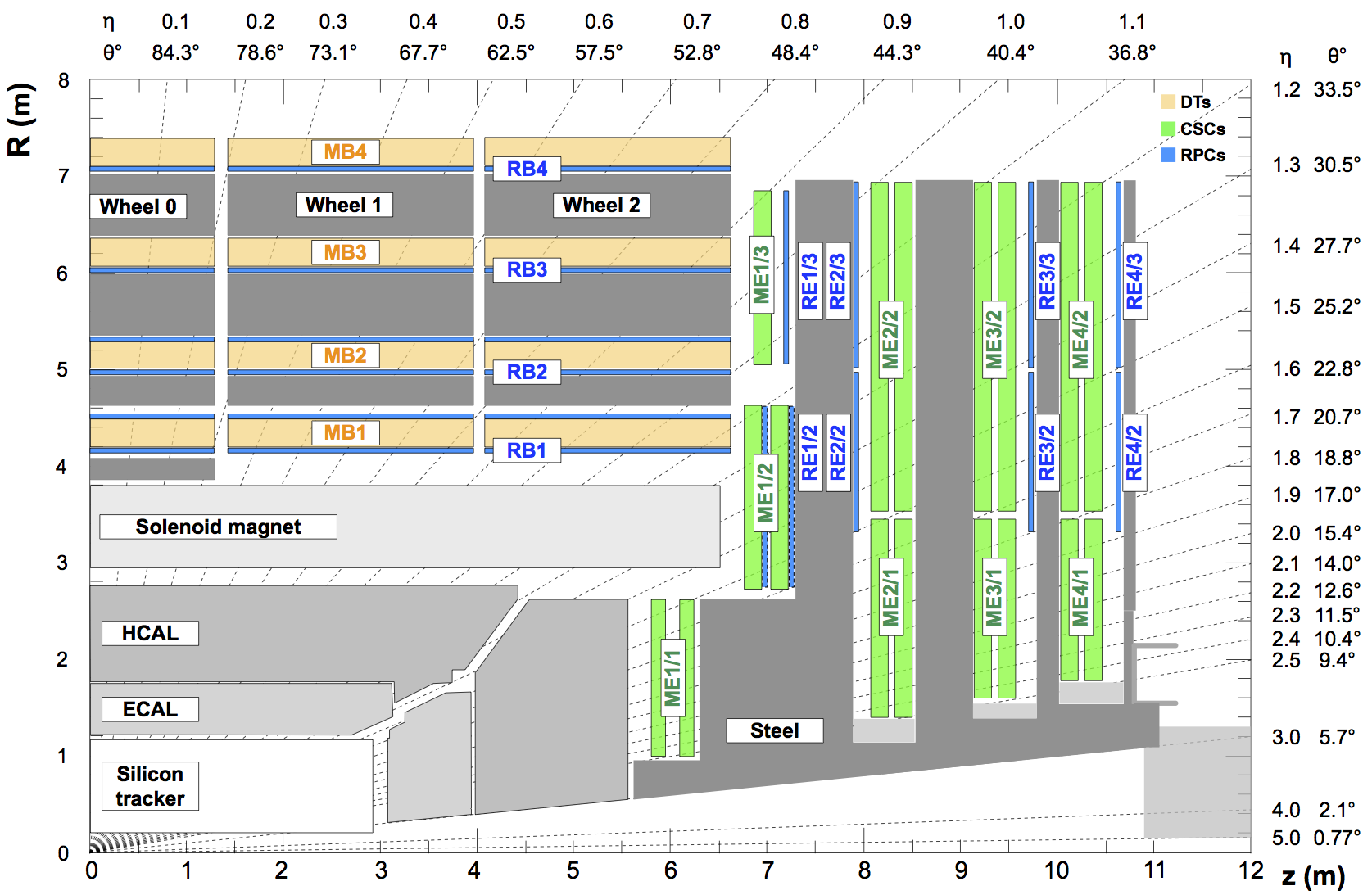}
    \caption{An \textit{R}-\textit{z} cross section of a quadrant of the CMS detector
      with the axis parallel to the beam (\textit{z}) running horizontally and the
      radius (\textit{R}) increasing upward.
      The interaction point is at the lower left corner.
      The locations of the various muon stations and the steel flux-return disks (dark
      areas) are shown.
      The drift tube stations (DTs) are labeled MB (``Muon Barrel'') and the cathode
      strip chambers (CSCs) are labeled ME (``Muon Endcap'').
      Resistive plate chambers (RPCs) are mounted in both the barrel and endcaps of
      CMS, where they are labeled RB and RE, respectively.
    }
    \label{fig:CMSquadrant}
  }
\end{figure}

A detailed description of the CMS detector, together with a definition of the
coordinate system and the relevant kinematic variables, can be found in
Ref.~\cite{CMSExp}.
A schematic diagram of the CMS detector is shown in Fig.~\ref{fig:CMSquadrant}.
The CMS detector has a cylindrical geometry that is azimuthally ($\phi$) symmetric
with respect to the beamline and features a superconducting magnet, which provides
a 3.8~T solenoidal field oriented along the beamline.
An inner tracker comprising a silicon pixel detector and a silicon strip tracker is
used to measure the momentum of charged particles in the pseudorapidity range
$\abs{\eta} < 2.5$.
The muon system is located outside the solenoid and covers the range
$\abs{\eta} < 2.4$.
It is composed of gaseous detectors sandwiched among the layers of the steel
flux-return yoke that allow a traversing muon to be detected at multiple points
along the track path.

\begin{figure}[ht]
  {
    \centering
    \includegraphics[width=1.0\textwidth]{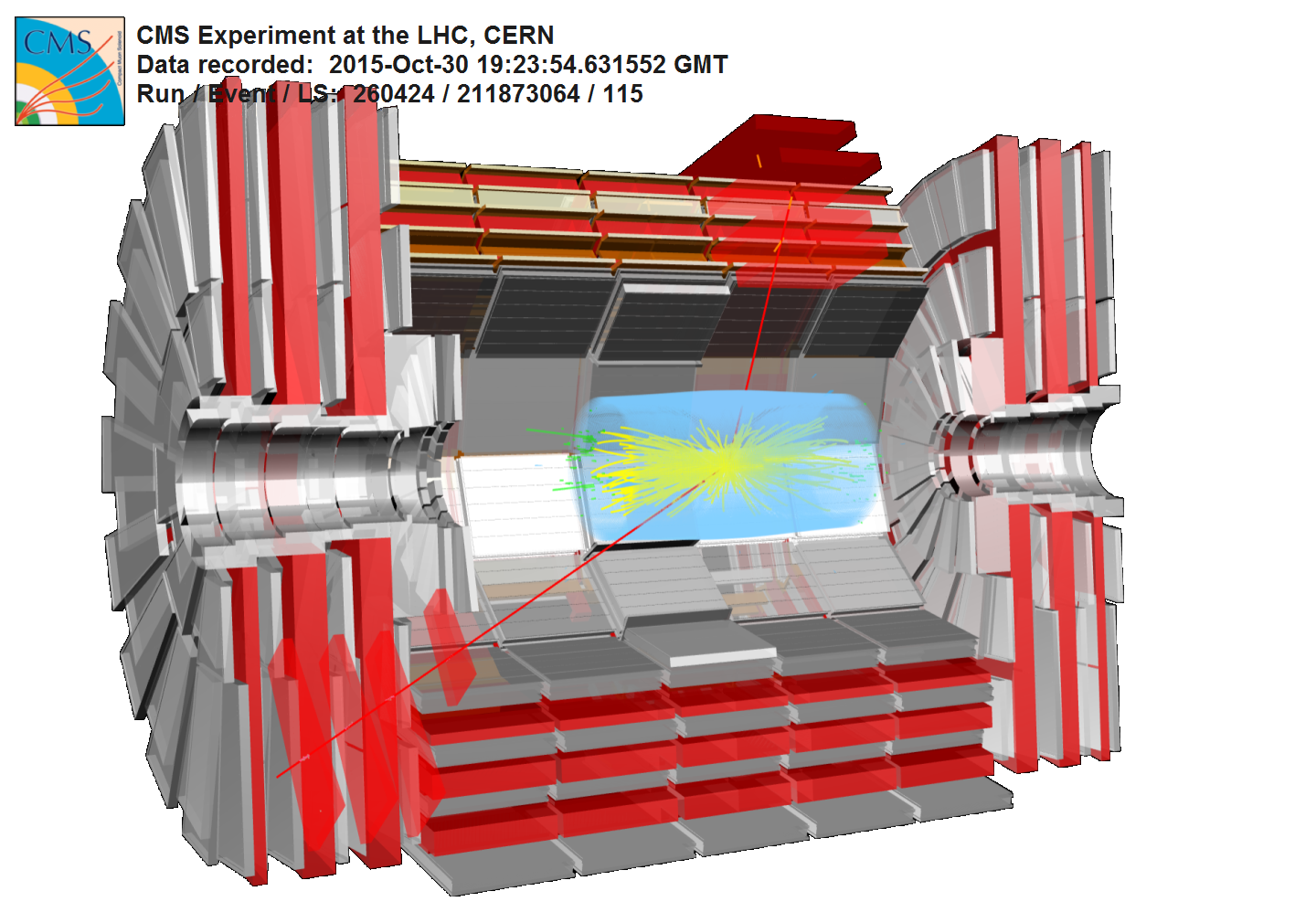}
    \caption{A {\Pp\Pp} collision event with two reconstructed muon tracks
      superimposed on a cutaway image of the CMS detector.
      The image has been rotated around the y axis, which makes the inner tracker
      appear offset relative to its true position in the center of the detector.
      The four layers of muon chambers are interleaved with three layers of the
      steel flux-return yoke.
      The reconstructed invariant mass of the muon pair is 2.4\TeV.
      One muon is reconstructed in the barrel with a transverse momentum (\pt)
      of 0.7\TeV, while the second muon is reconstructed in the endcap with \pt of
      1.0\TeV.
    }
    \label{fig:ppCollision_mumu}
  }
\end{figure}

Three types of gas ionization chambers were chosen to make up the CMS muon
system: drift tube chambers (DTs), cathode strip chambers (CSCs), and resistive
plate chambers (RPCs).
A detailed description of these chambers, including gas composition and
operating voltage, can be found in Ref.~\cite{DPGPerformance}.
The DTs are segmented into drift cells; the position of the muon is determined
by measuring the drift time to an anode wire of a cell with a shaped electric
field.
The CSCs operate as standard multi-wire proportional counters but add a finely
segmented cathode strip readout, which yields an accurate measurement of the
position of the bending plane (\textit{R}-$\phi$) coordinate at which the muon
crosses the gas volume.
The RPCs are double-gap chambers operated in avalanche mode and are primarily
designed to provide timing information for the muon trigger.
The DT and CSC chambers are located in the regions  $|\eta| < 1.2$ and
$0.9 < \abs{\eta} < 2.4$, respectively, and are complemented by RPCs in the range
$\abs{\eta} < 1.9$.
We distinguish three regions, naturally defined by the cylindrical geometry of
CMS, referred to as the barrel ($\abs{\eta} < 0.9$), overlap
($0.9 < |\abs{\eta}| < 1.2$), and endcap ($1.2 < \abs{\eta} < 2.4$) regions.
The chambers are arranged to maximize the coverage and to provide some overlap
where possible.
An event in which two muons are reconstructed, one in the barrel and one in the
endcap, is shown in Fig.~\ref{fig:ppCollision_mumu}.

In the barrel, a station is a ring of chambers assembled between two layers of the
steel flux-return yoke at approximately the same value of radius \textit{R}.
There are four DT and four RPC stations in the barrel, labeled MB1--MB4 and
RB1--RB4, respectively.
Each DT chamber consists of three ``superlayers'', each comprising four staggered
layers of parallel drift cells.
The wires in each layer are oriented so that two of the superlayers measure the muon
position in the bending plane (\textit{R}-$\phi$) and one superlayer measures the
position in the longitudinal plane (\textit{R}-$\theta$).
However, the chambers in MB4 have only the two \textit{R}-$\phi$ superlayers.
The two innermost RPC barrel stations, RB1 and RB2, are instrumented with two layers
of RPCs each, facing the innermost and outermost sides of the DT.
For stations 3 and 4 the RPCs have only one detection layer.
The RPC strips are oriented parallel to the wires of the DT chambers that measure the
coordinate in the bending plane.
From the readout point of view, every RPC is subdivided  into two or three $\eta$
partitions called ``rolls''~\cite{CMSTriggerTDR}.
Both DT and RPC barrel stations are arranged in five ``wheels'' along the \textit{z}
dimension, with 12 $\phi$-sectors per wheel.

In the endcap, a station is a ring of chambers assembled between two disks of the
steel flux-return yoke at approximately the same value of \textit{z}.
There are four CSC and four RPC stations in each endcap, labeled ME1--ME4 and
RE1--RE4, respectively.
Between Run~1 and Run~2, additional chambers were added in ME4 and RE4 to increase
redundancy, improve efficiency, and reduce misidentification rates.
Each CSC chamber consists of six staggered layers, each of which measures the muon
position in two coordinates.
The cathode strips are oriented radially to measure the muon position in the bending
plane (\textit{R}-$\phi$), whereas the anode wires provide a coarse measurement in
\textit{R}.
The RPC strips are oriented parallel to the CSC strips to measure the coordinate in
the bending plane, and each endcap chamber is divided into three $\abs{\eta}$ partitions
(rolls) identified by the  letters A, B, and C.
In the radial direction, stations are arranged in two or three "rings" of endcap RPCs
and CSCs.
In the inner rings of stations 2, 3, and 4, each CSC chamber subtends a $\phi$ angle
of 20\de; all other CSCs subtend an angle of 10\de.

\begin{table}[htbp]
  \topcaption{Properties and parameters of the CMS muon subsystems during the 2016
    data collection period.
  }
  \centering
  \begin{tabular}{l|c c c}
    \hline
    Muon subsystem                &    DT                               &    CSC                  & RPC              \\
    \hline
    $\abs{\eta}$ coverage         & 0.0--1.2                            & 0.9--2.4                & 0.0--1.9         \\
    Number of stations            & 4                                   & 4                       & 4                \\
    Number of chambers            & 250                                 & 540                     & Barrel: 480      \\
                                  &                                     &                         & Endcap: 576      \\
    Number of layers/chamber      & \textit{R}-$\phi$: 8; \textit{z}: 4 & 6                       & 2 in RB1 and RB2 \\
                                  &                                     &                         & 1 elsewhere      \\
    Number of readout channels    & 172 000                             & Strips: 266 112         & Barrel: 68 136   \\
                                  &                                     & Anode channels: 210 816 & Endcap: 55 296   \\
    Percentage of active channels &   98.4\%                            &   99.0\%                &  98.3\%          \\
    \hline
  \end{tabular}
  \label{tab:MuDetParameters}
\end{table}

Using these conventions, in this paper the performance of the DTs is specified
according to chamber type, labeled ``MBn$\pm$w'', where n is the barrel station
(increasing with \textit{R}), $+$ or $-$ specifies the \textit{z}-direction, and
w is the wheel (increasing with $\abs{z}$, with w = 0 centered at $z = 0$).
The CSCs are labeled ``ME$\pm$n/m'', where $+$ or $-$ specifies the
\textit{z}-direction, n is the endcap station (increasing with $\abs{z}$), and m
is the ring (increasing with \textit{R}).
If no sign is specified, the performance of the $+$ and $-$ stations are combined.
The inner ring of the CSC chambers in station 1 has a structure that is different
from the other rings; the primary difference is an additional division of ME1/1
into two $\eta$ partitions called a and b~\cite{DPGPerformance}.
An overview of the number of chambers per chamber type, number of readout channels,
and number of active channels in 2016 is given in Table~\ref{tab:MuDetParameters}.

The CMS trigger system consists of two stages~\cite{L1Paper} and is described in
more detail in Section~\ref{sec:trigger}.
A level-1 (L1) trigger based on custom-made electronics reduces the event rate
from 40\unit{MHz} (LHC bunch crossing rate) to a readout rate of 100\unit{kHz}.
For the muon component of the L1 trigger, CSC and DT chambers provide
``trigger primitives'' constructed from hit patterns consistent with muons that
originate from the collision region, and RPC chambers provide hit information.
When a specific bunch crossing is selected by the L1 algorithms as a potential
event, readout of the precision data from the CMS detector is initiated via
the ``L1-Accept'' (L1A) signal, which is synchronously distributed to all CMS
subsystems.
The high-level trigger (HLT), based on a farm of microprocessors, uses the
precision data to reconstruct events to further reduce the rate of data to
preserve for offline analysis to approximately 1\unit{kHz}.
Both L1 and HLT use information from the muon system to efficiently identify muons
over the broad energy range required for physics signatures of interest while
minimizing the trigger rate and operating within the available latency.

The LHC is a bunched machine, in which the accelerated protons are distributed in
bunches separated by one or more time steps of 25\unit{ns}.
The running conditions of the LHC have evolved continuously since the beginning of
its operation, and are expected to continue to evolve in the
future~\cite{Run1LHC,PapottiIPAC,WenningerIPAC}.
As a representative comparison, we compare the LHC conditions in fill 1440
(October 2010), included in the dataset analyzed in
Refs.~\cite{DPGPerformance,POGPerformance}, with the conditions in fill 5013 (June
2016), included in the 2016 data used in this paper.
Between these two fills, the center-of-mass energy increased from $\sqrt{s}=7\TeV$
to $\sqrt{s}=13\TeV$.
The maximum instantaneous luminosity increased by about a factor of 40, from
2\ten{32}\percms to 8\ten{33}\percms, as a result of the increases
in both the number of colliding bunches and the luminosity per bunch.
The number of colliding bunches increased by about a factor of 6, from 348 to 2028,
facilitated by the reduction of the spacing between proton bunches from
150\unit{ns} to 25\unit{ns}.
The average luminosity per bunch increased by about a factor of 6.5, from
0.6\ten{30}\percms to 3.9\ten{30}\percms, as a result of several
changes including increasing the number of protons per bunch, reducing the
transverse widths of the beams, and focusing the beams more
tightly~\cite{PapottiIPAC,WenningerIPAC}.
The combined increases in collision energy and luminosity per bunch caused the
average number of inelastic collisions per crossing (pileup) to increase by about
a factor of 8, from 3.6 to 28.

In order to prepare for these challenging LHC conditions and to exploit the
corresponding gain in luminosity, the CMS muon system was significantly modified
between Run~1 and Run~2.
As mentioned previously, additional RPC and CSC chambers, RE4 and ME4/2, were
installed in the fourth station to increase redundancy, improve efficiency, and
reduce misidentification rates.
The trigger and readout electronics were improved as part of the CMS-wide trigger
upgrade~\cite{CMSPhase1TriggerTDR}, including optical links in the DTs and CSCs
to increase bandwidth and to ease
maintenance~\cite{BMTFUpgrade,DTReadoutUpgrade}.
New electronics were installed in the CSC ME1/1 chambers to read out every
strip in the ME1/1a ring, covering $2.1 < \abs{\eta} < 2.4$.
These strips had been ganged together in Run~1, combining every 16th strip,
which led to a 3-fold ambiguity for the position of a hit on that strip plane.
The removal of the strip ganging in Run~2 leads to reduced capacitance, in turn
leading to reduced noise and a resulting improvement in the $\phi$ resolution
in ME1/1a.

\section{Muon reconstruction}
\label{sec:muonreco}

\subsection{Hit and segment reconstruction}
\label{subsec:localReco}

This section gives a brief overview of the ``local'' reconstruction algorithms
in the CMS muon detector.
Local reconstruction uses information from only a single muon chamber (RPC,
CSC, or DT) to specify the passage of a muon through the
chamber~\cite{DPGPerformance}.

Muons and other charged particles that traverse a muon subdetector ionize the
gas in the chambers, which eventually causes electric signals to be produced on
the wires and strips.
These signals are read out by electronics and are associated with well-defined
locations, generically called ``hits'', in the detector.
The precise location of each hit is reconstructed from the electronic signals
using different algorithms depending on the detector technology.

Hit reconstruction in a DT drift cell specifies the transverse distance between
the wire and the intersection of the muon trajectory with the plane containing
the wires in the layer.
The electrons produced through gas ionization by a muon crossing the cell are
collected at the anode wire.
A time-to-digital converter (TDC) registers their arrival time, $T_{\textsc{TDC}}$.
This time is then corrected by a time pedestal, $T_{\text{ped}}$, and multiplied
by the electron drift velocity, \textit{v}, to reconstruct the position of the
DT hit:
\begin{equation}
  \label{eq:DTHitPosition}
  \text{position} = (T_{\textsc{TDC}} - T_{\text{ped}}) \times v.
\end{equation}
The DT drift cell was designed to provide a uniform electric field so that the
drift velocity can be assumed to be mostly constant for tracks impinging on the
cell perpendicular to the plane of wires.
The effect of deviations from this assumption on the spatial resolution is
described in Section~\ref{sec:spatialresolution}.
In Equation~\ref{eq:DTHitPosition}, the time pedestal accounts for the time
from the bunch crossing until the trigger decision arrives at the chamber
electronics.
It includes the time-of-flight (at the speed of light) along a straight line
from the interaction region to the center of the wire, the average signal
propagation time along the wire, the generation of trigger primitives, the
processing by the L1 trigger electronics, the distribution of the L1A signals,
and the receipt of L1A back at the readout electronics on the chamber.
It also includes a wire-by-wire component that takes into account the
different signal paths within a chamber.
In another iteration, the time-of-flight and signal propagation time are
refined using the segment position from the orthogonal superlayer available
for MB1, MB2, and MB3.
The calibration of $T_{\text{ped}}$ and \textit{v} is described in detail in
Ref.~\cite{DPGPerformance}.
Effectively, the drift time is tuned to make $(T_{\textsc{TDC}}-T_{\text{ped}})=0$
for muons that cross the chamber at the location of the wire.

Hit reconstruction in a CSC layer measures the position of the traversing muon
by combining information from the cathode strips and anode wires.
The strips are radial, each subtending an angle of about 3\unit{mrad}
(different chamber types have different angular strip widths that range from
2.2 to 4.7\unit{mrad}) and can thus accurately measure the $\phi$ angle.
This is the bending direction of a muon traveling through the endcaps.
In the endcaps the solenoidal field is first parallel to the z direction but
then diverges radially, so a muon is first deflected in one azimuthal direction
and then deflected in the opposite direction, with the maximum deflection
occurring in the first station.
The wires are orthogonal to the strips, except in ME1/1 where they are tilted
to compensate for the Lorentz drift of ionization electrons in the
non-negligible magnetic field in this region.
They are ganged into wire groups of about 1--2\cm width, which results in
a coarser-grained measurement in the radial direction.
A CSC hit is reconstructed at the intersection points of hit strips and wire
groups.
A CSC reconstructed hit also has a measured time, which is calibrated such that
hits from muons produced promptly in the triggering bunch crossing have a time
distribution centered around zero.

Hit reconstruction in an RPC chamber requires clustering of hit strips.
A charged particle passing through the RPC produces an avalanche of
electrons in the gap between two plates.
This charge induces a signal on an external strip readout plane to identify
muons from collision events with a precision of a few\unit{ns}.
The strips are aligned with $\eta$ with up to 2\cm strip pitch, therefore
giving a few\cm spatial resolution in the $\phi$ coordinate.
Since the ionization charge from a muon can be shared by more than one strip,
adjacent strips are clustered to reconstruct one hit.
An RPC hit is reconstructed as the strip cluster centroid.

While the RPC chambers are single-layer chambers, the CSC and DT chambers are
multi-layer detectors where hits are reconstructed in each layer.
From the reconstructed hits, straight-line track ``segments'' are built within
each CSC or DT chamber.

Segment reconstruction in the DTs was modified prior to
Run~2~\cite{REF-DT-MEANTIMER}.
The calibration of $T_{\text{ped}}$ in Eq.~\ref{eq:DTHitPosition} implicitly
assumes that all muons take the same time to reach the reconstructed hit
position from the interaction region.
However, this assumption is not exactly true since hits could come from muons
originating from other bunch crossings (``out-of-time muons''), or could be
produced by heavy particles that travel at a reduced speed.
Any such shift in the muon crossing time would cause all hits produced within
a chamber to be shifted in space by the same amount.
Therefore, DT segment reconstruction was modified prior to Run~2 to include
time as third parameter, in addition to the intercept and slope of
the standard two-dimensional straight-line pattern recognition and fit
algorithm (in the plane transverse to the wire direction).
The inclusion of time into segment reconstruction allows spurious early hits,
produced by delta rays, to be removed from the segment reconstruction and
thus improves the spatial resolution (see Section~\ref{sec:spatialresolution}).
The segment time information is not needed in the muon track reconstruction
algorithm because of the negligible rate of accidentally matching out-of-time
segments.
The timing data are, however, kept with the reconstructed muon track
information to be used in physics analyses (see Section~\ref{sec:timing}).

\subsection{Muon track reconstruction}
\label{subsec:muonID}

In the standard CMS reconstruction procedure for {\Pp\Pp}
collisions~\cite{TRK-11-001,POGPerformance,CRAFT}, tracks are first
reconstructed independently in the inner tracker (tracker track) and in the
muon system (standalone-muon track), and then used as input for muon track
reconstruction.

Tracker tracks are built using an iterative approach, running a sequence of
tracking algorithms, each with slightly different logic.
After each iteration step, hits that have been associated with reconstructed
tracks are removed from the set of input hits to be used in the following
step.
This approach maintains high performance and reduces processing
time~\cite{TRK-11-001}.

\textit{Standalone-muon tracks} are built by exploiting information from muon
subdetectors to gather all CSC, DT, and RPC information along a muon
trajectory using a Kalman-filter technique~\cite{Fruhwirth:1987fm}.
Reconstruction starts from seeds made up of groups of DT or CSC segments.

\textit{Tracker muon tracks} are built ``inside-out'' by propagating tracker
tracks to the muon system with loose matching to DT or CSC segments.
Each tracker track with transverse momentum $\pt > 0.5\GeV$ and a total
momentum $p > 2.5\GeV$ is extrapolated to the muon system.
If at least one muon segment matches the extrapolated track, the tracker track
qualifies as a tracker muon track.
The track-to-segment matching is performed in a local \textit{(x,y)} coordinate
system defined in a plane transverse to the beam axis, where \textit{x} is the
better-measured coordinate (in the \textit{R}-$\phi$ plane) and \textit{y} is
the coordinate orthogonal to it.
The extrapolated track and the segment are matched either if the absolute value
of the difference between their positions in the \textit{x} coordinate is
smaller than 3\cm, or if the ratio of this distance to its uncertainty (pull) is
smaller than 4.

\textit{Global muon tracks} are built ``outside-in'' by matching standalone-muon
tracks with tracker tracks.
The matching is done by comparing parameters of the two tracks propagated onto
a common surface.
A combined fit is performed with the Kalman filter using information from both
the tracker track and standalone-muon track.

Owing to the high efficiency of the tracker track and muon segment
reconstruction, about 99\% of the muons produced within the geometrical
acceptance of the muon system are reconstructed either as a global muon track
or as a tracker muon track, and very often as both.
Global muons and tracker muons that share the same tracker track are merged
into a single candidate.

Tracker muons have high efficiency in regions of the CMS detector with less
instrumentation (for routing of detector services) and for muons with low \pt.
The tracker muons that are not global muons typically match only to segments
in the innermost muon station, but not other stations.
This increases the probability of muon misidentification since hadron shower
remnants can reach this innermost muon station (punch-through).
Global muon reconstruction, which uses standalone-muon tracks, is
designed to have high efficiency for muons penetrating through more than one
muon station, which reduces the muon misidentification rate compared to
tracker muons.
By fully exploiting the information from both the inner tracker and the muon
system, the \pt measurement of global muons is also improved compared to
tracker muons, especially for $\pt > 200\GeV$.
Muons reconstructed only as standalone-muon tracks have worse momentum
resolution and a higher admixture of cosmic muons than global or tracker muons.

Reconstructed muons are fed into the CMS particle flow (PF)
algorithm~\cite{PRF-14-001}.
The algorithm combines information from all CMS subdetectors to identify and
reconstruct all individual particles for each event, including electrons,
neutral hadrons, charged hadrons, and muons.
For muons, PF applies a set of selection criteria to candidates reconstructed
with the standalone, global, or tracker muon algorithms.
The requirements are based on various quality parameters from the muon
reconstruction (described in Section~\ref{subsec:muonid}), as well as make
use of information from other CMS subdetectors (\eg, isolation as described in
Section~\ref{subsec:isolation}).

Prior to Run~2, two muon-specific calculations were added to the tracker track
reconstruction to keep reconstruction and identification efficiency as high as
possible under high-pileup conditions~\cite{PRF-14-001}.
In the first calculation, tracker tracks identified as tracker muons are
rebuilt by relaxing some quality constraints to increase track hit efficiency.
In the second, standalone-muon tracks with $\pt > 10\GeV$ that fulfill a
minimal set of quality requirements are used to seed an outside-in inner
tracking reconstruction step.
This additional set of tracks is combined with those provided by the inner
tracking system and is exploited to build global and tracker muons.

\subsection{Muon identification}
\label{subsec:muonid}

A set of variables was studied and selection criteria were defined to allow
each analysis to tune the desired balance between efficiency and purity.
Some variables are based on muon reconstruction, such as track fit $\chi^2$,
the number of hits per track (either in the inner tracker or in the muon
system, or both), or the degree of matching between tracker tracks and
standalone-muon tracks (for global muons).
The muon segment compatibility is computed by propagating the tracker track to
the muon system, and evaluating both the number of matched segments in all
stations and the closeness of the matching in position and
direction~\cite{CRAFT}.
The algorithm returns values in a range between 0 and 1, with 1 representing
the highest degree of compatibility.
A kink-finding algorithm splits the tracker track into two separate tracks at
several places along the trajectory.
For each split the algorithm makes a comparison between the two separate
tracks, with a large $\chi^2$ indicating that the two tracks are incompatible
with being a single track.
Other variables exploit inputs from outside the reconstructed muon track, such
as compatibility with the primary vertex (the reconstructed vertex with the
largest value of summed physics-object $\pt^2$~\cite{Phase2TDR}).
Using these variables, the main identification types of muons used in CMS
physics analyses include:

\begin{itemize}

\item \textit{Loose muon identification (ID)} aims to identify prompt muons
  originating at the primary vertex, and muons from light and heavy flavor
  decays, as well as maintain a low rate of the misidentification of charged
  hadrons as muons.
  A loose muon is a muon selected by the PF algorithm that is also either a
  tracker or a global muon.

\item \textit{Medium muon ID} is optimized for prompt muons and for muons from
  heavy flavor decay.
  A medium muon is a loose muon with a tracker track that uses hits from more
  than 80\% of the inner tracker layers it traverses.
  If the muon is only reconstructed as a tracker muon, the muon segment
  compatibility must be greater than 0.451.
  If the muon is reconstructed as both a tracker muon and a global muon, the
  muon segment compatibility need only be greater than 0.303, but then the
  global fit is required to have goodness-of-fit per degree of freedom
  (\chisqdof) less than 3, the position match between the tracker muon and
  standalone-muon must have $\chi^2 < 12$, and the maximum $\chi^2$ computed
  by the kink-finding algorithm must be less than 20.
  The constraints on the segment compatibility were tuned after application of
  the other constraints to target an overall efficiency of 99.5\% for muons
  from simulated {\PW} and {\cPZ} events.

\item \textit{Tight muon ID} aims to suppress muons from decay in flight and
  from hadronic punch-through.
  A tight muon is a loose muon with a tracker track that uses hits from at
  least six layers of the inner tracker including at least one pixel hit.
  The muon must be reconstructed as both a tracker muon and a global muon.
  The tracker muon must have segment matching in at least two of the muon
  stations.
  The global muon fit must have $\chisqdof < 10$ and include at least one hit
  from the muon system.
  A tight muon must be compatible with the primary vertex, having a transverse
  impact parameter $\abs{dXY} < 0.2\cm$ and a longitudinal impact parameter
  $\abs{dz} < 0.5\cm$.

\item \textit{Soft muon ID} is optimized for low-\pt muons for \PB-physics and
  quarkonia analyses.
  A soft muon is a tracker muon with a tracker track that satisfies a high
  purity flag~\cite{TRK-11-001} and uses hits from at least six layers of the
  inner tracker including at least one pixel hit.
  The tracker muon reconstruction must have tight segment matching, having
  pulls less than 3 both in local \textit{x} and in local \textit{y}.
  A soft muon is loosely compatible with the primary vertex, having
  $\abs{dXY} < 0.3\cm$ and $\abs{dz} < 20\cm$.

\item \textit{High momentum muon ID} is optimized for muons with
  $\pt > 200\GeV$.
  A high momentum muon is reconstructed as both a tracker muon and a global
  muon.
  The requirements on the tracker track, the tracker muon, and the transverse
  and longitudinal impact parameters are the same as for a tight muon, as
  well as the requirement that there be at least one hit from the muon system
  for the global muon.
  However, in contrast to the tight muon, the requirement on the global muon
  fit {\chisqdof} is removed.
  The removal of the $\chi^2$ requirement prevents inefficiencies at high \pt
  when muons radiate large electromagnetic showers as they pass through the
  steel flux-return yoke, giving rise to additional hits in the muon chambers.
  A requirement on the relative \pt uncertainty, $\sigma(\pt)/\pt < 30\%$, is
  used to ensure a proper momentum measurement.

\end{itemize}

\subsection{Determination of muon momentum}

The default algorithm used by CMS to determine the muon momentum is the
\textit{Tune-P} algorithm~\cite{POGPerformance}.
For each muon, the \textit{Tune-P} algorithm selects the \pt measurement from
one of the following refits based on goodness-of-fit information and
$\sigma(\pt)/\pt$ criteria to reduce tails in the momentum resolution
distribution due to poor quality fits.

\begin{itemize}
\item \textit{Inner-Track fit} determines the momentum using only information
  from the inner tracker.
  While various fit methods are used to add information from the muon detector
  to improve the measurement of the momentum at high \pt, for muons with
  $\pt < 200\GeV$, the contribution from the muon system to the momentum
  measurement is marginal.
  Therefore, the inner-track fit is highly favored by \textit{Tune-P} at
  low momentum.

\item \textit{Tracker-Plus-First-Muon-Station fit} starts with the hits
  from the global muon track and performs a refit using only information from
  the inner tracker and the innermost muon station containing hits.
  The innermost station provides the best information about momentum within
  the muon system.

\item \textit{Picky fit} aims at properly determining the momentum for events
  in which showering occurred within a chamber.
  This algorithm again starts with the hits from the global muon track, but
  in chambers that have a large hit occupancy (\ie likely from a shower) the
  refit uses only the hits that are compatible with the extrapolated
  trajectory (based on $\chi^2$).

\item \textit{Dynamic-Truncation fit} accounts for cases when energy losses
  cause significant bending of the muon trajectory.
  The algorithm propagates the tracker track to the innermost station and
  performs a refit adding hits from the segment closest to the extrapolated
  trajectory, if compatible.
  Starting from the refit, the algorithm is repeated for each station
  propagating outward.
  If no compatible hit is found in two consecutive muon stations, the
  algorithm stops.

\end{itemize}

The \textit{Tune-P} algorithm was validated using cosmic ray muons, muons from
{\Pp\Pp} collisions, and Monte Carlo simulations generated using different
misalignment scenarios.
Both the core and the tails of the momentum, curvature, and invariant mass
distributions were studied to ensure that no significant biases in the muon
momentum assignment are introduced by the algorithm.

The PF algorithm refines the information from \textit{Tune-P}, exploiting
information from the full event, by selecting refits that significantly improve
the balance of missing \pt and by using a post-processing algorithm designed to
preserve events that contain genuine missing energy~\cite{PRF-14-001}.
The PF momentum assignment was also validated using Monte Carlo simulation and
muons from {\Pp\Pp} collisions.

\subsection{Muon isolation}
\label{subsec:isolation}

To distinguish between prompt muons and those from weak decays within jets,
the isolation of a muon is evaluated relative to its \pt by summing up the
energy in geometrical cones,
$\Delta R = \sqrt{\smash[b]{(\Delta\phi)^2 + (\Delta\eta)^2}}$, surrounding the
muon.
One strategy sums reconstructed tracks (track based isolation), while another
uses charged hadrons and neutral particles coming from PF (PF isolation).

For the computation of PF isolation~\cite{PRF-14-001}, the \pt of charged
hadrons within the $\Delta R$ cone originating from the primary vertex are
summed together with the energy sum of all neutral particles (hadrons and
photons) in the cone.
The contribution from pileup to the neutral particles is corrected by
computing the sum of charged hadron deposits originating from pileup
vertices, scaling it by a factor of 0.5, and subtracting this from the
neutral hadron and photon sums to give the corrected energy sum from neutral
particles.
The factor of 0.5 is estimated from simulations to be approximately the
ratio of neutral particle to charged hadron production in inelastic
proton-proton collisions.
The corrected energy sum from neutral particles is limited to be positive or
zero.

For both strategies, tight and loose working points are defined to achieve
efficiencies of 95\% and 98\%, respectively.
They are tuned using simulated tight muons from {\Zmm} decays with
$\pt > 20\GeV$.
The values for the tight and loose working points for PF isolation within
$\Delta R < 0.4$ are 0.15 and 0.25, respectively, while the values for
track based isolation within $\Delta R < 0.3$ are 0.05 and 0.10.
The efficiency of the working points to reject muons in jets was tested in
simulated multi-jet QCD events (events comprised uniquely of jets produced
through the strong interaction) and simulated events containing a {\PW} boson
plus one or more jets (\PW+jets).

\section{Data and simulated samples}
\label{sec:samples}

Results shown in this paper come from one of two data sets:  approximately
2\fbinv of {\Pp\Pp} collisions collected in 2015, which will be called
``2015 data'', and approximately 4\fbinv of {\Pp\Pp} collisions collected in
2016, which will be called ``2016 data''.
The data set that was used for each result in this paper was chosen depending
on the availability of the data and the analyst.
In any case, the results represent the CMS muon performance in Run 2 no matter
which data set is used, since the peak luminosity delivered by LHC in 2015 and
2016 differed only by about a factor of three, which is small compared to the
factor of 40 difference between 2010 and 2016 as described in
Section~\ref{sec:detectors}.
The selected data samples consist of events with a pair of reconstructed muons
with low \pt thresholds.
Further event criteria are applied depending on the analyses performed, and are
described in detail later.

The performance results most directly applicable to physics analyses are
presented in this paper using the 2015 data.
These data are compared with simulations from several Monte Carlo event
generators for signal and background processes.
The Drell-Yan $\text{\cPZ}/\text{\PGg}^* \to \text{\Pl}^{+}\text{\Pl}^{-}$ signal
sample is generated at next-to-leading order (NLO) with
\MGvATNLO 2.3.3~\cite{Alwall:2014hca}.
The background samples of \PW+jets and of \ttbar pairs with one or more
jets (\ttbar+jets) are also produced with the same generator.
The background from single top quark {\cPqt\PW} production is
generated at NLO with \POWHEG v1.0~\cite{Re:2010bp}.
The \PYTHIA 8.212~\cite{REF-PYTHIA6P4,REF-PYTHIA8P2} package is used for QCD
events enriched in muon decays, parton showering, hadronization, and
simulation of the underlying event via tune
\textsc{CUETP8M1}~\cite{underlyingEvent}, using \textsc{NNPDF}2.3
LO~\cite{REF-NNPDF} as the default set of parton distribution functions.
For all processes, the detector response is simulated using a detailed
description of the CMS detector based on the \GEANTfour
package~\cite{REF-GEANT} and event reconstruction is performed with the same
algorithms as used for the data.
The simulated samples include pileup, and the events are weighted so that the
pileup distribution matches the 2015 data, having an average pileup of about
11.

\section{Spatial resolution}
\label{sec:spatialresolution}

The spatial resolution of a muon subdetector is quantified by the width of the
distribution of residuals between the reconstructed and expected hit positions.
The expected position is estimated from the segment fit.
The resolution is obtained from the residual width by applying standard analytical
factors calculated from the ``hat matrix'' that relates the residuals from a fit to
the fitted measurements, and hence the widths of the residual distributions to the
intrinsic resolution of the measurements~\cite{Rencher}.
These factors differ for CSC, in which the reconstructed hit used for the residual
is excluded from the segment fit, and for DT where the hit is included.
Both CSCs and DTs are designed to make a precise measurement in the direction of
bending of a muon track because this directly affects the measurement of the
momentum.
This is the azimuthal direction, measured in the CSCs by the strips, and in the DTs
by the $\phi$ superlayers.

\begin{figure}[ht]
  {
    \centering
    \includegraphics[width=1.0\linewidth]{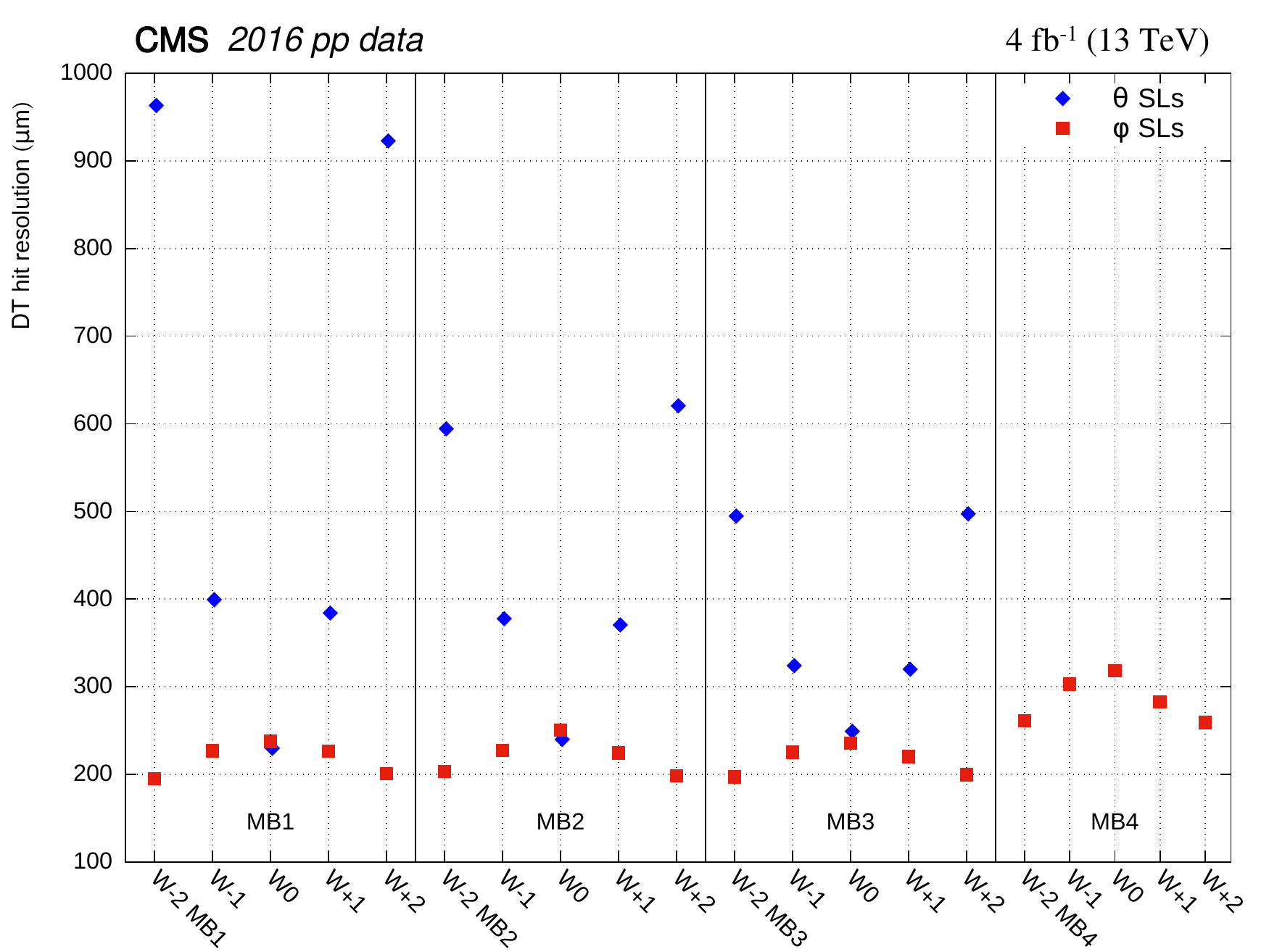}
    \caption{Reconstructed hit resolution for DT $\phi$ superlayers (squares) and
      DT $\theta$ superlayers (diamonds) measured with the 2016 data, plotted as
      a function of station and wheel.
      The uncertainties in these values are smaller than the marker size in the
      figure.
    }
    \label{fig:dpg3}
  }
\end{figure}

The spatial resolution of the DTs is determined by computing the value of the
residual for each hit used to reconstruct each segment.
Typically, eight residual values are computed for each $\phi$ segment and four
for each $\theta$ segment.
Due to the azimuthal symmetry of the DTs, a single residual distribution is filled
with all hits having the same wire orientation from all chambers in the same wheel
and station.
The width of each residual distribution is converted to position resolution using
the standard analytically computed factors described above.
Figure~\ref{fig:dpg3} shows the spatial resolution of DT hits sorted by station,
wheel, and wire orientation.
The resolution in the $\phi$ superlayers (\ie, in the bending plane) is better
than 250\mum in MB1, MB2 and MB3, and better than 300\mum in MB4.
In the $\theta$ superlayers, the resolution varies from about 250 to 600\mum
except in the outer wheels of MB1.

Within every station, both $\theta$ and $\phi$ superlayers show symmetric behavior
with respect to the $z = 0$ plane, as expected from the detector symmetry.
In wheel 0, where tracks from the interaction region are mostly perpendicular to all
layers, the resolution is the same for $\theta$ and $\phi$ superlayers.
From wheel 0 toward the forward region, tracks from the interaction region have
increasing values of $\abs{\eta}$; this affects $\theta$ and $\phi$ superlayers in
opposite ways.
In the $\theta$ superlayers the increasing inclination angle degrades the linearity
of the distance-drift time relation, thus worsening the resolution.
In contrast, in $\phi$ superlayers the inclination angle increases the track path
within the tube (along the wire direction), thus increasing the ionization charge
and improving the resolution.
The resolution of the $\phi$ superlayers is worse in MB4 because no $\theta$
measurement is available, so no corrections can be applied to account for the muon
time of flight and the signal propagation time along the wire.
The DT spatial resolution in the 2016 data is improved by about 10\% compared to
the 2010 results~\cite{DPGPerformance} as a result of the improved track
reconstruction method in Run~2 that removes spurious early hits, produced by delta
rays, from the segment reconstruction (see Section~\ref{subsec:localReco}).

The spatial resolution of the CSCs is studied using locally reconstructed segments that
have exactly one hit per layer.
For each segment, the hit in one layer is dropped and the segment is re-fitted with the
remaining five hits.
The residual between the dropped hit and the new fit is calculated as $R\,\Delta\phi$
in the \textit{R}-$\phi$ plane, which is the precision coordinate measured by the
strips and the direction of the magnetic bending of the muon.
This procedure is repeated for each layer.
These residuals are approximately Gaussian and the residual widths are converted to
position resolution by using standard analytical factors~\cite{Rencher}.
The spatial resolution of the CSC strip measurement depends on the relative position at
which a muon crosses a strip: it is better for a muon crossing near a strip edge than at
the center because then more of the induced charge is shared between that strip and its
neighbor, allowing a better estimate of the center of the charge distribution.
To benefit from this fact, alternate layers in a CSC are staggered by half a strip width,
except in the ME1/1 chambers where the strips are narrower and the effect is small.
Resolutions are measured separately for the central half of a strip width
($\sigma_{\textsc{C}}$) and the quarter strip-width at each edge
($\sigma_{\textsc{E}}$)~\cite{DPGPerformance}.
The layer measurements are combined to give an overall resolution $\sigma$ per CSC
station by $1/\sigma_{\text{station}}^2 = 6/\sigma_{\text{layer}}^2$ (ME1/1 chambers) and
$1/\sigma_{\text{station}}^2 = 3/\sigma_{\textsc{C}}^2 + 3/\sigma_{\textsc{E}}^2$ (chambers
other than ME1/1).
Table~\ref{Tab:dpg2} summarizes the mean  spatial resolution in each CSC station and
ring.
The design specifications for the spatial resolutions in the CSC system were 75\mum for
ME1/1 chambers and 150\mum for the others.
These resolutions were chosen so that the contribution of the chamber spatial resolution
to the muon momentum resolution is less than or comparable to the contribution of multiple
scattering.

\begin{table}
  \topcaption{CSC transverse spatial resolution per station (6 hits) measured
    for all chamber types with 2016 data, compared to those measured in
    2015 and 2012.
  }
  \centering
  \begin{tabular}{lccc}
    \hline
                 & \multicolumn{3}{c}{Spatial resolution ($\mu\mathrm{m}$)} \\
    Station/ring & Run~1       &      \multicolumn{2}{c}{Run~2}             \\
                 & 2012        & 2015          & 2016                       \\
    \hline
    ME1/1a       &  66         &  48           &  45                     \\
    ME1/1b       &  57         &  54           &  52                     \\
    ME1/2        &  93         &  93           &  90                     \\
    ME1/3        & 108         & 110           & 105                     \\
    ME2/1        & 132         & 130           & 125                     \\
    ME2/2        & 140         & 142           & 134                     \\
    ME3/1        & 125         & 125           & 120                     \\
    ME3/2        & 142         & 143           & 135                     \\
    ME4/1        & 127         & 128           & 123                     \\
    ME4/2        & 147         & 143           & 134                     \\
    \hline
  \end{tabular}
  \label{Tab:dpg2}
\end{table}

The precision of the CSC measurements is dominated by systematic effects, and the
statistical uncertainties arising from the fits to the residual distributions are
small ($<$0.2\%).
The precision is controlled by the size of the induced charge distribution on the
strip plane, which is affected by geometry (the width of the strips), gas gain
(high voltage, gas mix, gas pressure), and sample selection (momenta, angle of
incidence).
The gas mix and high voltage are stringently maintained constant during CSC
operation, and muon samples are selected to be as close as possible for the
purposes of these comparisons.
The CSCs operate at atmospheric pressure, but a decrease of atmospheric pressure
of 1\% increases the gas gain by approximately 7\%, so the values in the table
have all been normalized to 965\unit{mbar}, a value typical of the annual
average atmospheric pressure at CMS.
In this manner we obtain reproducible resolutions typically within 1--2\mum, as
can be seen from the values in Tab.~\ref{Tab:dpg2} for the columns for 2012 and
2015 (other than for ME1/1a).
The approximately 25\% improvement in resolution in ME1/1a CSCs between 2012 and
later is because of the removal of the strip ganging that was used in the first
CMS running periods.
The improved resolution is not directly related to the spatial nature of the
ganging---every 16th strip was ganged into a single channel, rather than
combining neighboring strips.
Instead, the improvement is because of the reduction of capacitance, and hence
noise, with the removal of this ganging.
The spatial resolution values for 2016 are systematically better than expected,
and this was eventually traced to an incorrectly calibrated gas flowmeter that
led to a slightly increased argon fraction in the gas mix in early 2016.
Once this was corrected\footnote{Better spatial resolution is not the only
  consideration in choice of gas mix for CSC operation in CMS. The gas mix is
  just one of many parameters of the system design that were optimized to
  provide the required spatial resolution while maintaining stable and robust
  operation of the detectior and maximum longevity of the chambers in the LHC
  environment.}
the measured values returned to those seen in earlier running periods.

The spatial resolution of the RPCs is studied by extrapolating segments from the
closest CSC or DT to the plane of strips in the chamber under study.
The residuals are calculated transverse to the direction of the strips, which is
also the direction of the bending of muons in the magnetic field.
The residual is defined as the transverse distance between the center of the
reconstructed RPC cluster and the point of intersection of the extrapolated
segment with the plane of strips.
For each station and layer, a residual distribution is filled and fit with a
Gaussian.
The $\sigma$ parameter of these fits varies between 0.78--1.27\cm in the barrel and
0.89--1.38\cm in the endcap.
These values are compatible with the resolution expected from the widths of the
strips and are consistent with the 2010 results~\cite{DPGPerformance}.

The spatial compatibility between tracker tracks, reconstructed with the inner tracker,
and segments, reconstructed in the muon chambers, is of primary importance and is
extensively used in the muon ID criteria presented in Section~\ref{subsec:muonID}.
The residuals between extrapolated tracker tracks and segments are studied using the
tag-and-probe technique~\cite{POGPerformance}.
Oppositely charged dimuon pairs are selected from a sample collected with a single-muon
trigger.
The tag is a tight muon with tight PF isolation, which is geometrically matched with the
trigger ($\Delta R < 0.1$ between the tracker track and the 4-vector reconstructed by HLT).
The probe is a tracker muon, which passes track-based isolation and tracker track quality
requirements, that is propagated to each of the DT or CSC chambers it traverses.
The segment matching in the definition of a tracker muon is loose enough not to bias this
measurement.

\begin{figure}[h!]
  {
    \centering
   \includegraphics[width=0.42\textwidth]{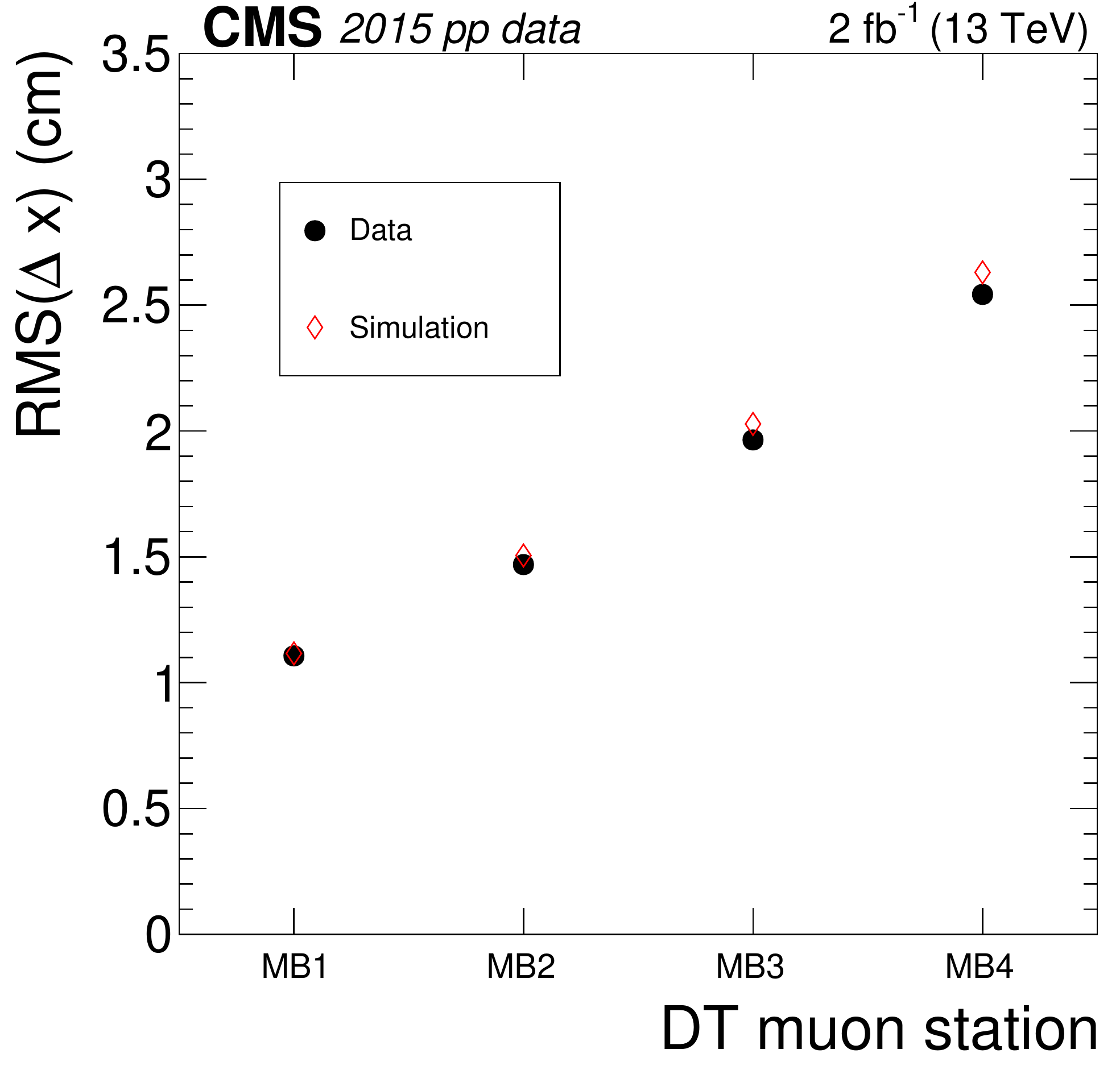}
   \includegraphics[width=0.42\textwidth]{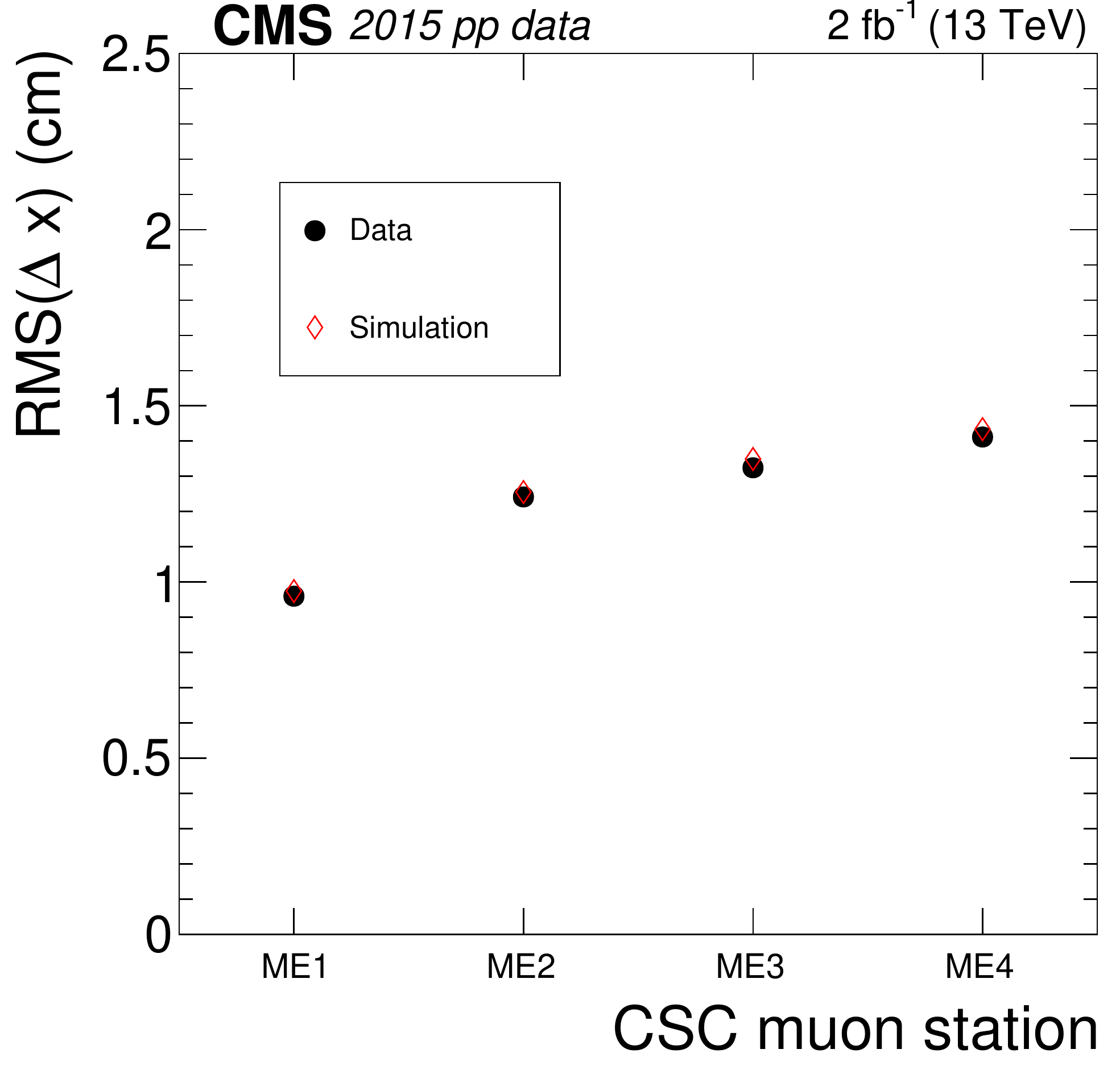}

   \includegraphics[width=0.28\textwidth]{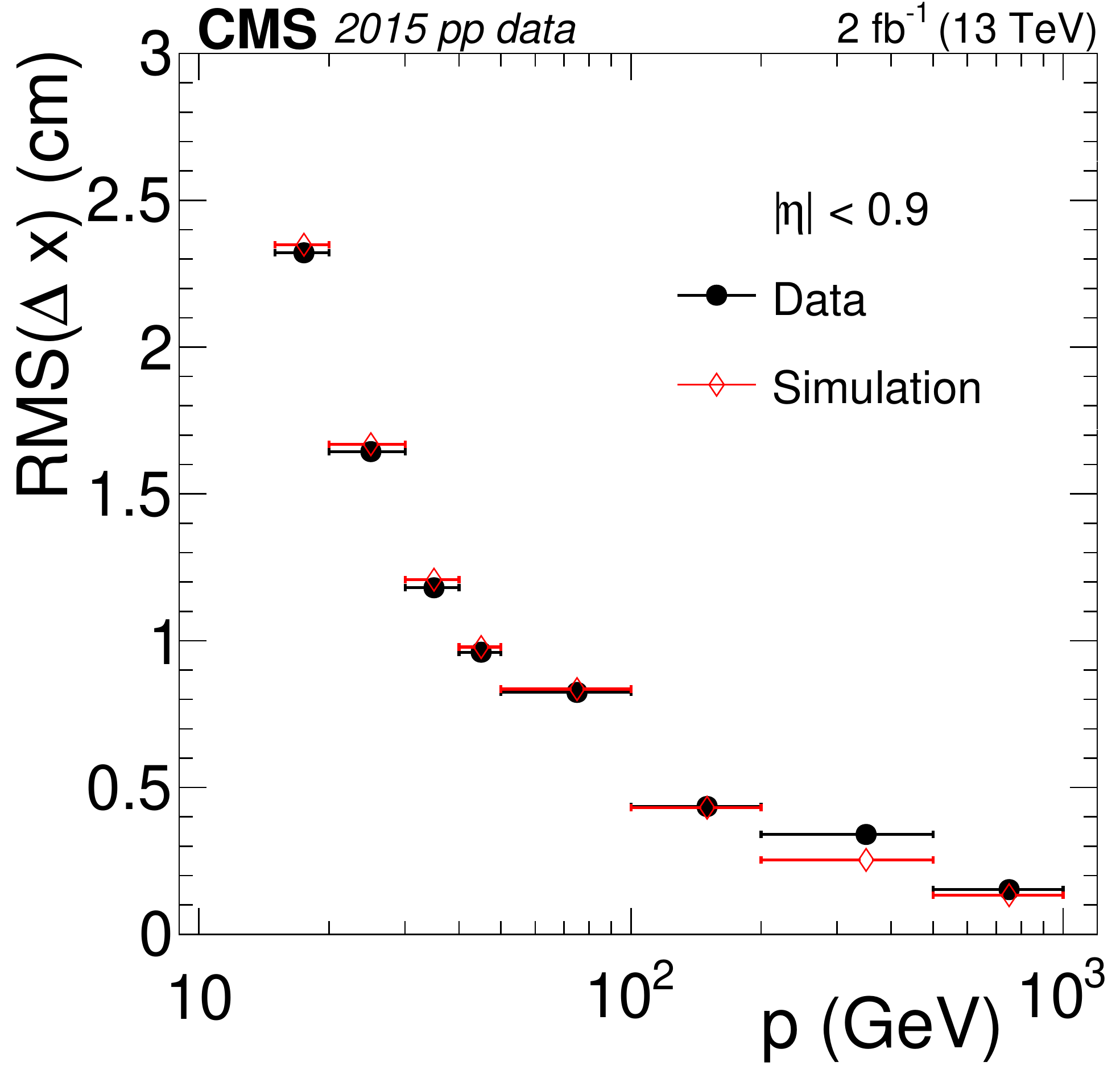}
   \includegraphics[width=0.28\textwidth]{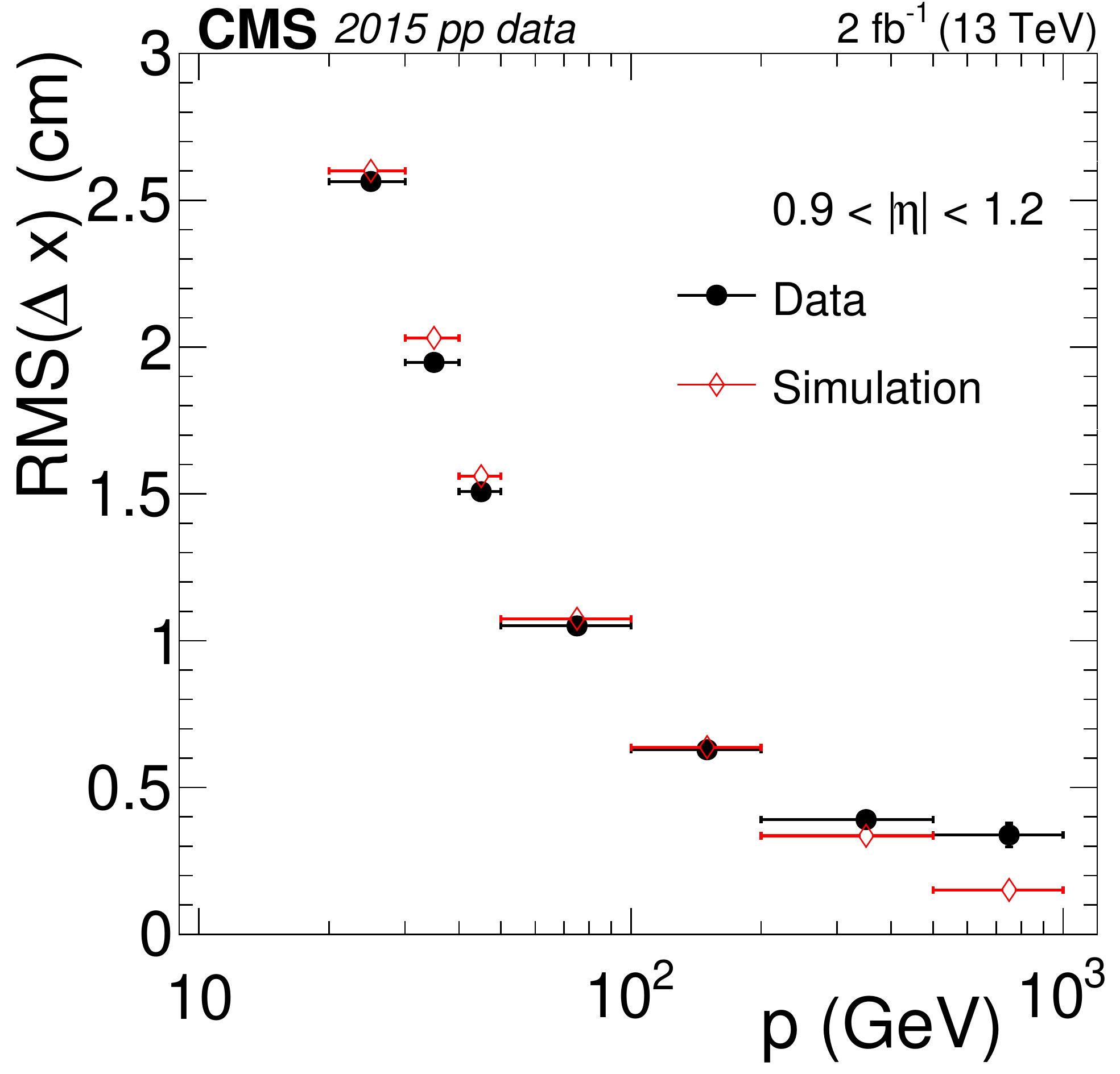}
   \includegraphics[width=0.28\textwidth]{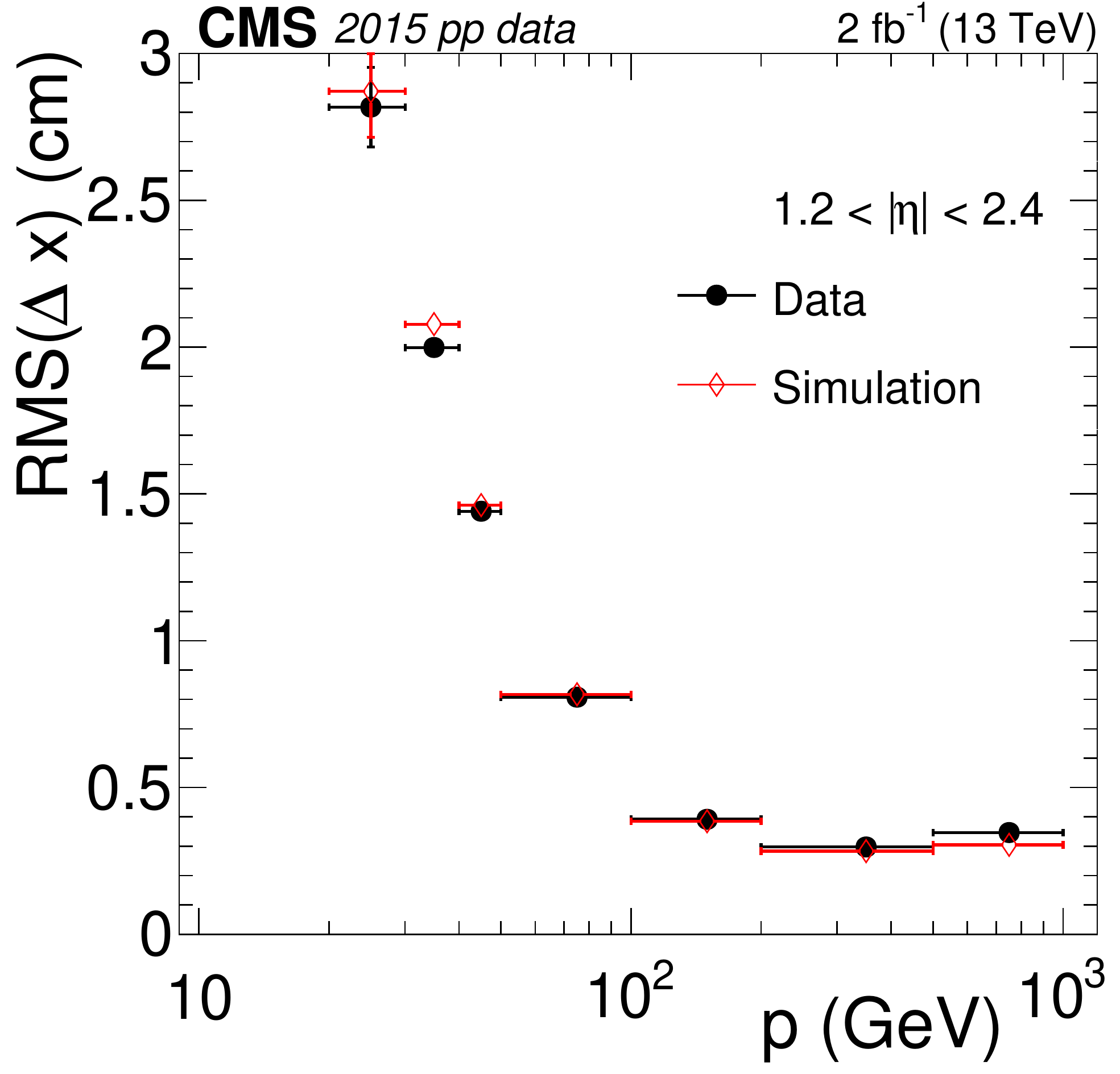}
    \caption{The RMS of transverse residuals between reconstructed segments and propagated
      tracker tracks, measured in 2015 data.  Results are plotted as a function of:
      (upper left) MB station in the DTs;
      (upper right) ME station in the CSCs;
      (lower left) momentum $p$ in station 1 of the barrel region ($\abs{\eta} \le 0.9$);
      (lower center) momentum $p$ in station 1 of the overlap region ($0.9 \le \abs{\eta} \le 1.2$);
      (lower right) momentum $p$ in station 1 of the endcap region ($1.2 \le \abs{\eta} \le 2.4$).
      The vertical error bars represent the statistical uncertainties of the RMS, and are
      smaller than the marker size for most data points.
    }
    \label{fig:MuonResiduals}
  }
\end{figure}

The transverse residual, $\Delta x$, is computed in the chamber local reference frame
for the coordinate measuring the muon position in the bending plane ($\phi$).
It corresponds to the distance between the position of the propagated tracker track and the
segment in the chamber.
The RMS of the distribution of $\Delta x$ is shown in Fig.~\ref{fig:MuonResiduals} for
2015 data and simulated $\text{\cPZ}/\text{\PGg}^* \to \text{\Pl}^{+}\text{\Pl}^{-}$ decays.
There is reasonable agreement between the data and simulation.
The alignment precision of the data (using the techniques described in
Ref.~\cite{TrackerAlignment} with the full 2015 data set) and of the simulation
(corresponding to what would be obtained with about 1\fbinv of data) is about 100--200\mum,
and thus is not a dominant effect in these results.
Figures~\ref{fig:MuonResiduals}a and \ref{fig:MuonResiduals}b show the RMS as a function
of station for DT and CSC chambers, respectively.
The RMS increases as the muon station number increases, which is expected because of the
larger amount of material traversed by the muons and the resultant multiple scattering.
The RMS of the residual evaluated in the first muon station is shown as a function of
momentum in Fig.~\ref{fig:MuonResiduals}c and Fig.~\ref{fig:MuonResiduals}e in the
barrel and endcap regions, respectively, while Fig.~\ref{fig:MuonResiduals}d shows the
overlap region between the two.
The RMS decreases with momentum because of the reduction in multiple scattering.
The spatial resolution in Fig.~\ref{fig:MuonResiduals} is not directly comparable with
the results in Ref.~\cite{POGPerformance} because the analysis used on the 2015 data
reduced the contamination from muons that do not come from the primary interaction.

\section{Efficiency}
\label{sec:efficiency}

\subsection{Hit and segment efficiency}
\label{subsec:hiteff}

The hit reconstruction efficiency is calculated as the ratio of the number of
reconstructed hits divided by the number of expected hits.
The measurement provided by the detecting unit under study is excluded from the
computation of the expected hit position.

\begin{figure}[h!]
  {
    \centering
    \includegraphics[width=0.42\textwidth]{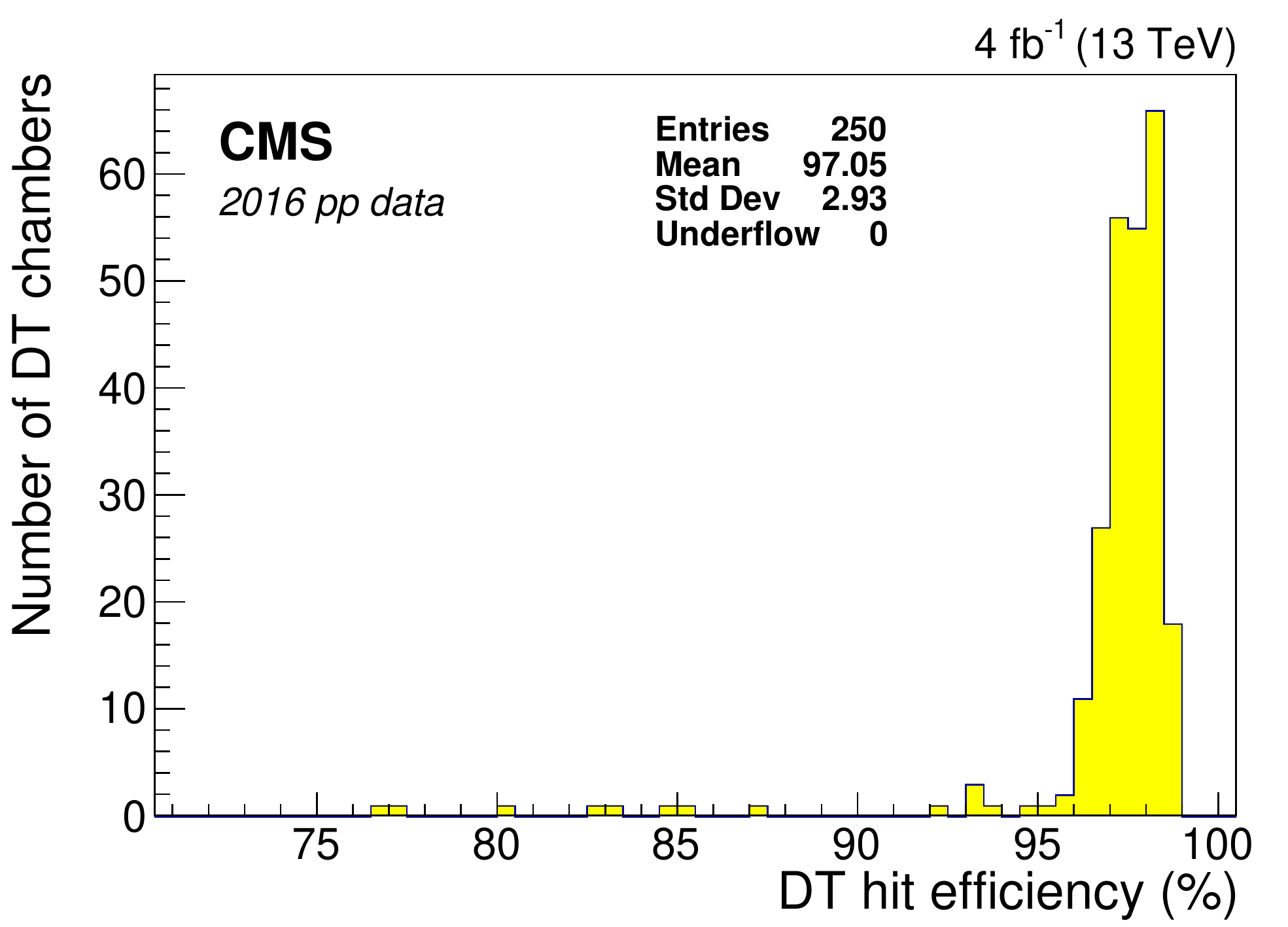}
    \includegraphics[width=0.42\textwidth]{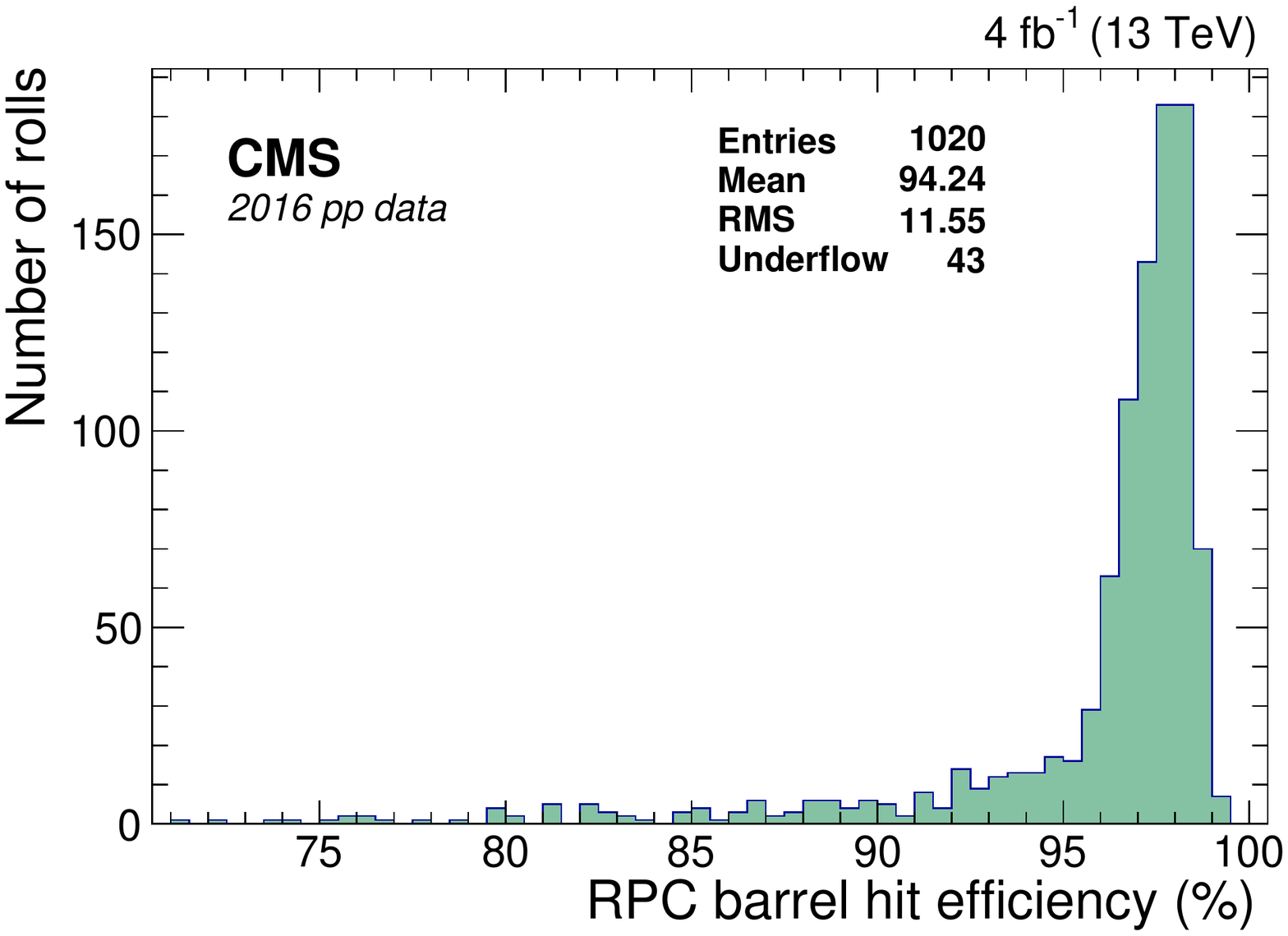}

    \includegraphics[width=0.42\textwidth]{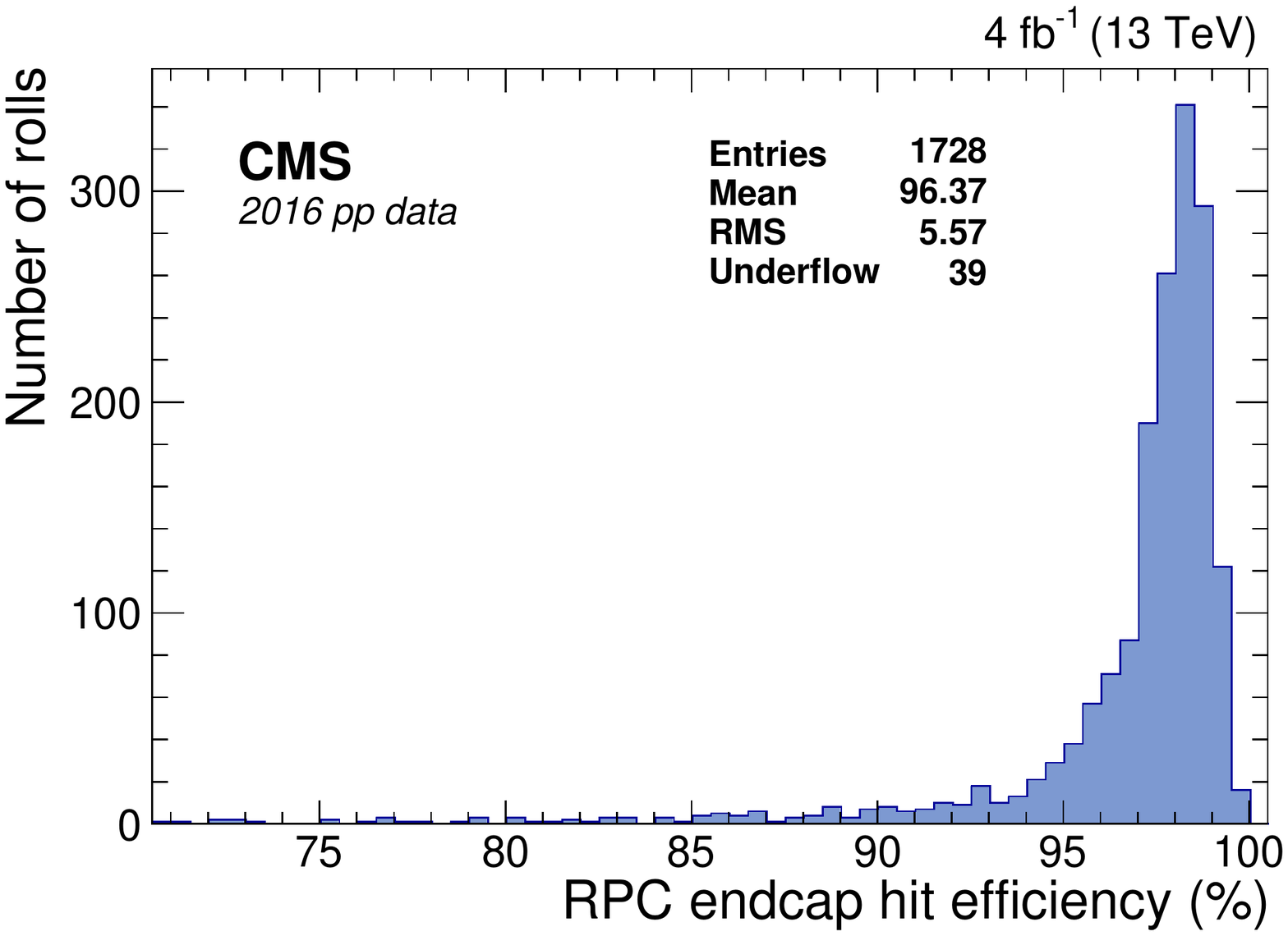}
    \caption{Hit reconstruction efficiency measured with the 2016 data in (upper left) DT,
      (upper right) RPC barrel, and (lower) RPC endcap chambers.
    }
    \label{fig:dpg1}
  }
\end{figure}

The hit reconstruction efficiency of the DTs is studied using segments.
To ensure high quality segment reconstruction, segments are required to have at
least one reconstructed hit in all layers except the layer under study.
For the efficiency of a $\phi$ layer, this implies that the $\phi$ segment must
have at least seven associated hits, while for $\theta$ layers, the $\theta$
segment must have at least three associated hits.
In addition to the high quality of the segment in the view under study, there
must be a segment constructed in both $\phi$ and $\theta$ views to ensure the
presence of a genuine muon crossing the chamber.
For $\phi$ superlayers, backgrounds are reduced by requiring the segment
inclination to be smaller than 45\de (by construction, muons from the
interaction region are mostly orthogonal to the wire plane).
The intersection of this segment with the layer under study determines the
position of the expected hit within a specific tube and increments the
denominator in the efficiency calculation.
The numerator is incremented if a hit is reconstructed in this tube.
The distribution of the hit reconstruction efficiency for each DT chamber is
shown in Fig.~\ref{fig:dpg1}a.
The average value of the DT hit reconstruction efficiency is 97.1\% including the
dead cells reported in Table~\ref{sec:intro}.
The average efficiency in the 2016 data is consistent with the 2010
average~\cite{DPGPerformance} within 1\%.

The hit reconstruction efficiency of the RPCs is studied with a tag-and-probe
technique.
Muon pairs are selected from an event sample collected with a single-muon
trigger.
The tag is a tight muon that is geometrically matched with the trigger
($\Delta R < 0.1$ between the tracker track and the 4-vector reconstructed by HLT).
The probe is a tracker muon matched to a DT or CSC segment that is extrapolated
to RPC chambers.
For each RPC roll that the extrapolated probe traverses, the denominator in the
efficiency calculation is incremented and a matching hit is sought.
The numerator is incremented if the absolute value of the difference between the
hit position and the extrapolated probe position is smaller than 10\cm, or if the
ratio of this distance to its uncertainty (pull), including the extrapolation
uncertainty, is less than 4.
Figures~\ref{fig:dpg1}b and c show the efficiency for all RPC barrel and endcap
rolls, respectively.
The average hit efficiency is 94.2\% for the RPC barrel and 96.4\% for the RPC
endcaps, with negligible accidental contributions from noise.
The underflow entries are from rolls with efficiency lower than 70\% caused by
known hardware problems:  chambers with gas leaks in the barrel, and low voltage
problems in the endcap.
The rolls with zero efficiency (Tab.~\ref{tab:MuDetParameters}) are included in
the underflow and the average efficiency.
Results on RPC hit efficiency from 2010~\cite{DPGPerformance} and 2016 are
consistent within 1\%.

\begin{figure}[ht]
  {
    \centering
    \includegraphics[width=0.8\linewidth]{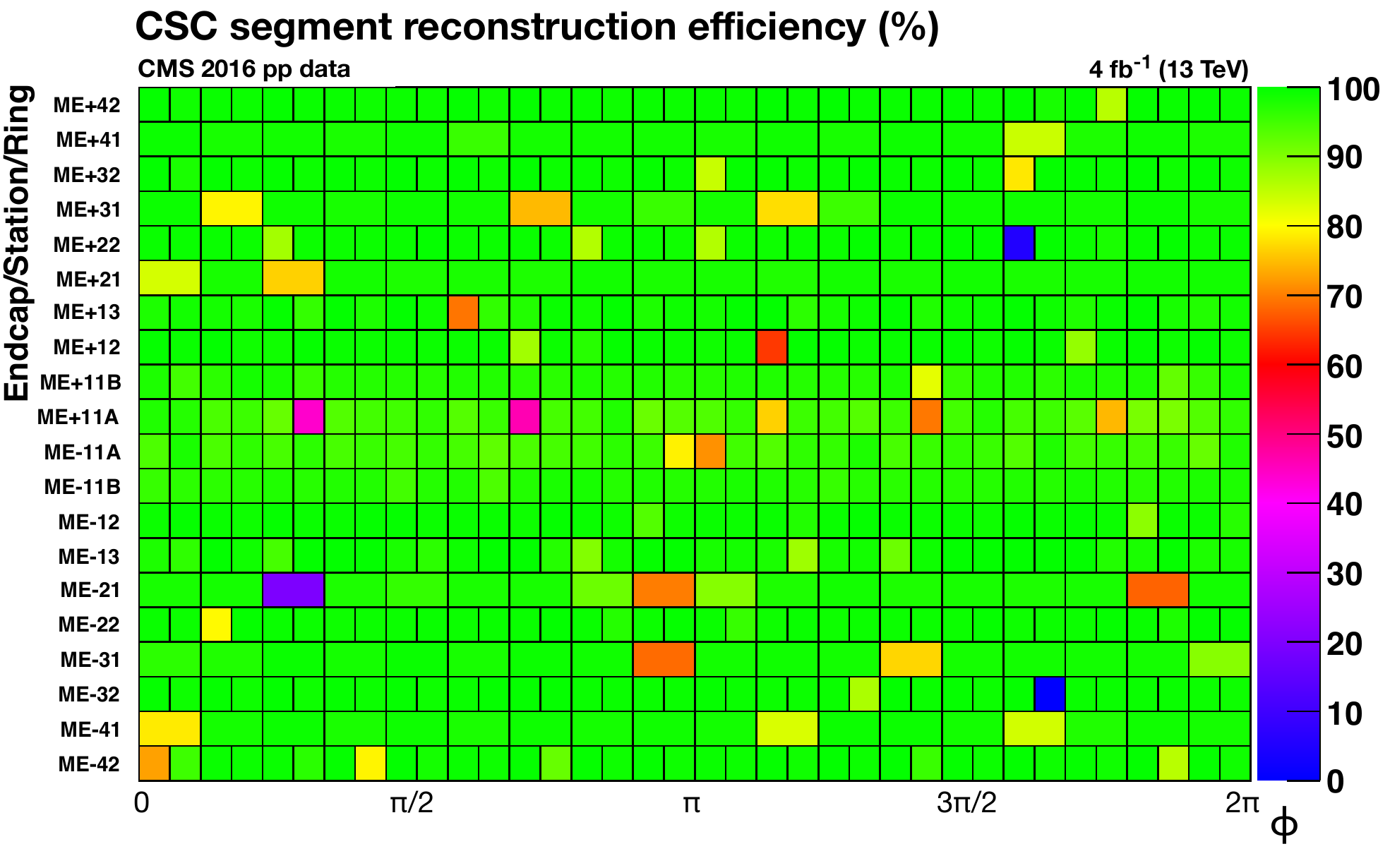}
    \caption{The efficiency (in percent) of each CSC in the CMS endcap muon detector
      to provide a locally reconstructed track segment as measured from 2016 data.
    }
    \label{fig:dpg5}
  }
\end{figure}

Muons rarely fail to traverse an entire CSC so the CSC readout
system ~\cite{CMSMuonTDR} requires hits compatible with a charged track crossing
a chamber, which suppresses readout of hits from several sources of
uninteresting background.
In order to read out a cathode front-end board, which services 16 strip channels
in each of the six layers of a CSC, the basic pattern of hits expected for a CSC
trigger primitive must occur in coincidence with a level-1 trigger from CMS.
A trigger primitive requires at least 4 layers in a CSC containing strip hits,
with a pattern consistent with those created by muons originating at the
{\Pp\Pp} collision point.
This readout suppression complicates the interpretation of straightforward
measurements of CSC layer-by-layer hit efficiencies, but since the muon track
reconstruction uses segments, and not individual hits, it is the segment
efficiency that is most important to system operation.
This can be directly measured using the tag-and-probe method.
The tag is required to be a tracker muon and the probe is a tracker track that
is projected to the muon system.
To reduce background and ensure that the probe actually enters the chamber under
consideration, compatible hits are also required in a downstream CSC.
In the case of station 4, an upstream segment is required.
Figure~\ref{fig:dpg5} shows a summary map of the measured reconstructed segment
efficiency for each CSC.
The average CSC segment reconstruction efficiency is 97.4\%.
A few of the 540 chambers have known inefficiencies, usually caused by one or
more faulty electronics boards that cannot be repaired without major intervention
requiring the dismantling of the system.
There are also occasional temporary failures of electronics boards that last for
a few hours or days and can be recovered without major intervention.
Both contribute to a reduced segment efficiency in a localized region.
The average CSC segment efficiency in the 2016 data is within 1\% of that
observed in 2010~\cite{DPGPerformance}.

\subsection{Reconstruction, identification and isolation efficiency}
\label{subsec:recoeff}

The efficiency for muons is studied with the tag-and-probe method beginning with
tracker tracks as probes.
The value of the efficiency is computed by factorizing it into several
components~\cite{POGPerformance}:
\begin{equation}
  \label{eqn:efficiency}
  \epsilon_\mu = \epsilon_{\text{track}} \times \epsilon_{\text{reco+ID}} \times
  \epsilon_{\text{iso}} \times \epsilon_{\text{trig}}.
\end{equation}
Each component of $\epsilon_\mu$ is determined individually.
The efficiency of the tracker track reconstruction is
$\epsilon_{\text{track}}$~\cite{TRK-11-001}.
The reconstruction+ID efficiency, $\epsilon_{\text{reco+ID}}$, contains both the
efficiency of muon reconstruction in the muon system, including the matching of
this muon to the tracker track, and the efficiency of the ID criteria.
The efficiency of muon isolation, $\epsilon_{\text{iso}}$, is studied relative to
a probe that has passed the specified muon ID.
The efficiency of the trigger, $\epsilon_{\text{trig}}$, is described in detail in
Section~\ref{sec:muontriggerreco}.
The application of Eq.~\ref{eqn:efficiency} is dependent on the specific needs of
each analysis.
For example, if an analysis does not require isolation, $\epsilon_{\text{iso}}$ is
removed from the equation and $\epsilon_{\text{trig}}$ is computed relative to
reconstructed muons without an isolation requirement.

As described in Ref.~\cite{POGPerformance}, the combinatorial background of
tag-probe pairs not coming from the {\cPZ} resonance (where the probe is usually a
charged hadron misidentified as a muon) is subtracted by performing a simultaneous
fit to the invariant mass spectra for passing and failing probes with identical
signal shape and appropriate background shapes; the efficiency is then computed
from the normalizations of the signal shapes in the two spectra.
Given the high multiplicity of tracks in proton-proton collision events, using a
tracker track as the probe leads to a high combinatorial background in low-\pt
bins, which can result in large uncertainties in the background subtraction
method.
To mitigate this effect, the efficiency measurement is performed using only
the tag-and-probe pairs for which a single probe is associated with the tag.
The same method is also applied to simulated {\Zmm} events.

\begin{figure}[hbt!]
  {
    \centering
    \includegraphics[width=0.45\textwidth]{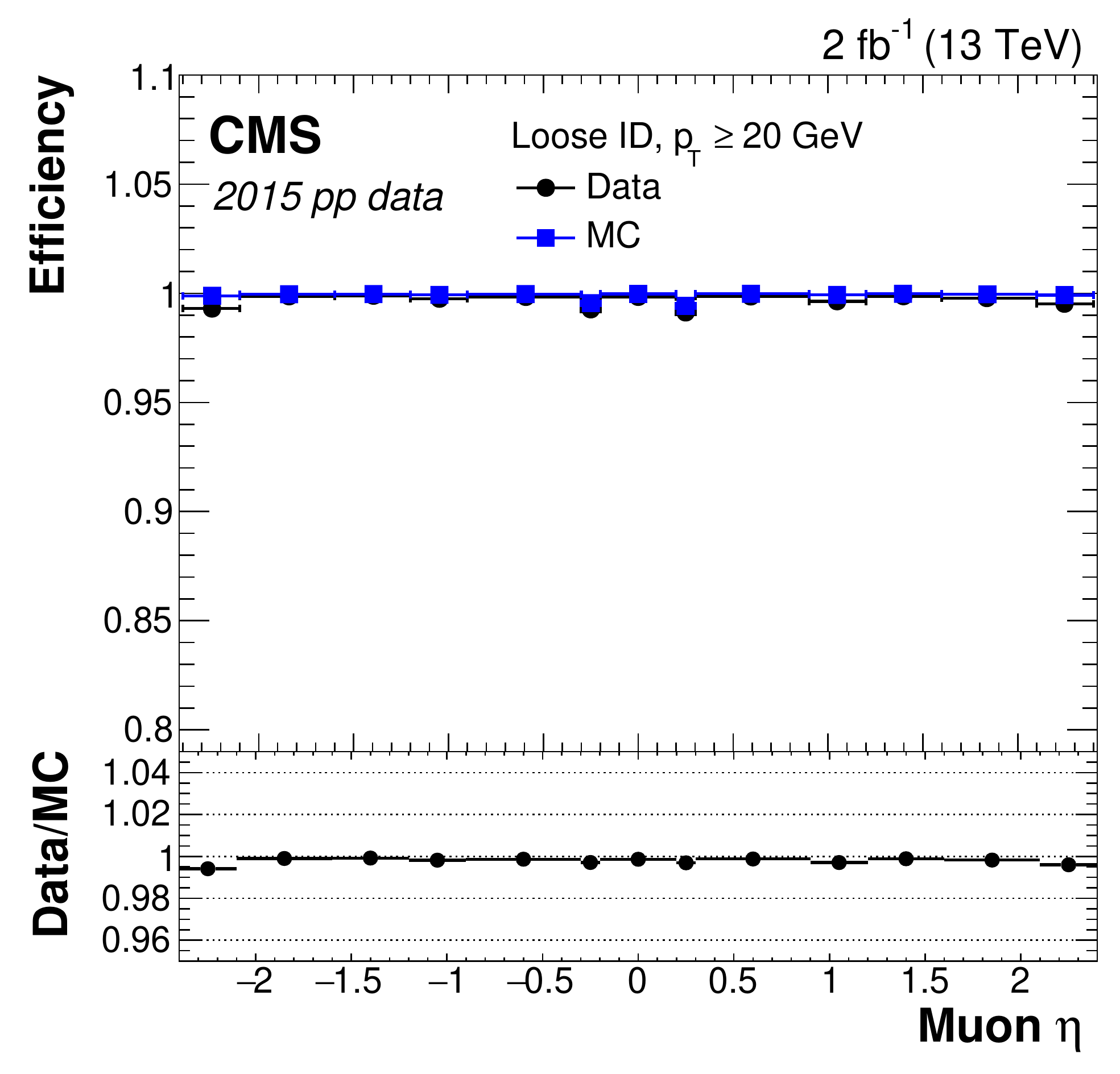} \hspace*{1.0cm}
    \includegraphics[width=0.45\textwidth]{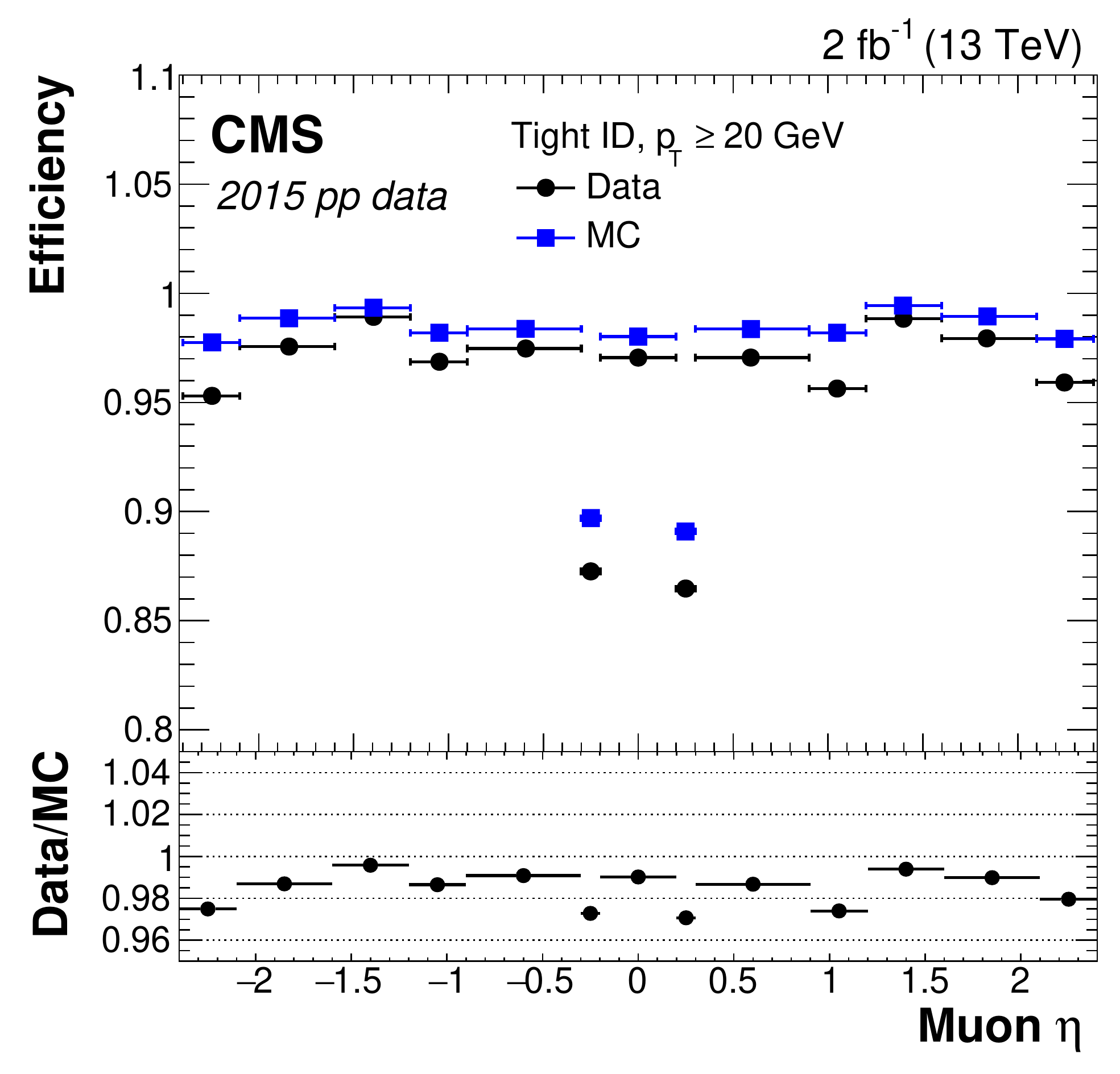}
    \caption{Tag-and-probe efficiency for muon reconstruction and identification
      in 2015 data (circles), simulation (squares), and the ratio (bottom inset)
      for loose (left) and tight (right) muons with $\pt > 20\GeV$.
      The statistical uncertainties are smaller than the symbols used to display
      the measurements.
    }
    \label{fig:TightIDefficiency}
  }
\end{figure}

The $\epsilon_{\text{reco+ID}}$ for loose muons and for tight muons are shown as a
function of $\eta$ in Fig.~\ref{fig:TightIDefficiency}, for both data and
simulation.
The loose ID efficiency exceeds 99\% over the entire $\eta$ range, and the
data and simulation agree to within 1\%.
As a function of \pt between 20\GeV and 200\GeV (where the efficiency is
measured with reasonably small uncertainty), the loose ID efficiency is constant
with fluctuations well within 1\%.
The tight ID efficiency varies between 95\% and 99\%, depending on $\eta$, and the
data and simulation agree to within 1--3\%.
The dips in efficiency close to $\abs{\eta} = 0.3$ are due to the regions with less
instrumentation between the central muon wheel and the two neighboring wheels.
In Fig.~\ref{fig:TightIDefficiency}b, the simulation is systematically higher than
the data as a result of small imperfections in the model, which are revealed by
the stringent requirements for a muon to satisfy tight ID criteria.
In the endcap, differences between the data and simulation arise when the muon is
required to be global with a combined fit that has valid hits in the muon system,
whereas in the barrel segment matching and global reconstruction contribute to the
discrepancy in a similar way.
Tracker track quality constraints contribute to a discrepancy of less than 0.5\%
over the full $\eta$ range.

A hadron may be misidentified as a prompt muon if the hadron decays in flight, or
if hadron shower remnants penetrate through the calorimeters and reach the muon
system (punch-through), or if there is a random matching between a hadron track in
the inner tracker and a segment or standalone-muon in the muon system.
The probability of hadrons to be misidentified as muons is measured by using data
samples of pions and kaons from resonant particle decays collected with jet
triggers~\cite{POGPerformance}.
The probability of pions to be misidentified as loose muons in both data and
simulation is about 0.2\% while for tight muons it is about 0.1\%.
In the same way, 0.5\% of kaons are misidentified as loose muons and 0.3\% as
tight muons in both data and simulation.
The uncertainty in these measurements is at the level of 0.05\% and is dominated
by the limited statistical precision.
Within uncertainties, the misidentification probabilities are independent of \pt.
These results are in good agreement with Run~1.

\begin{figure}[hbt!]
  {
    \centering
   \includegraphics[width=0.45\textwidth]{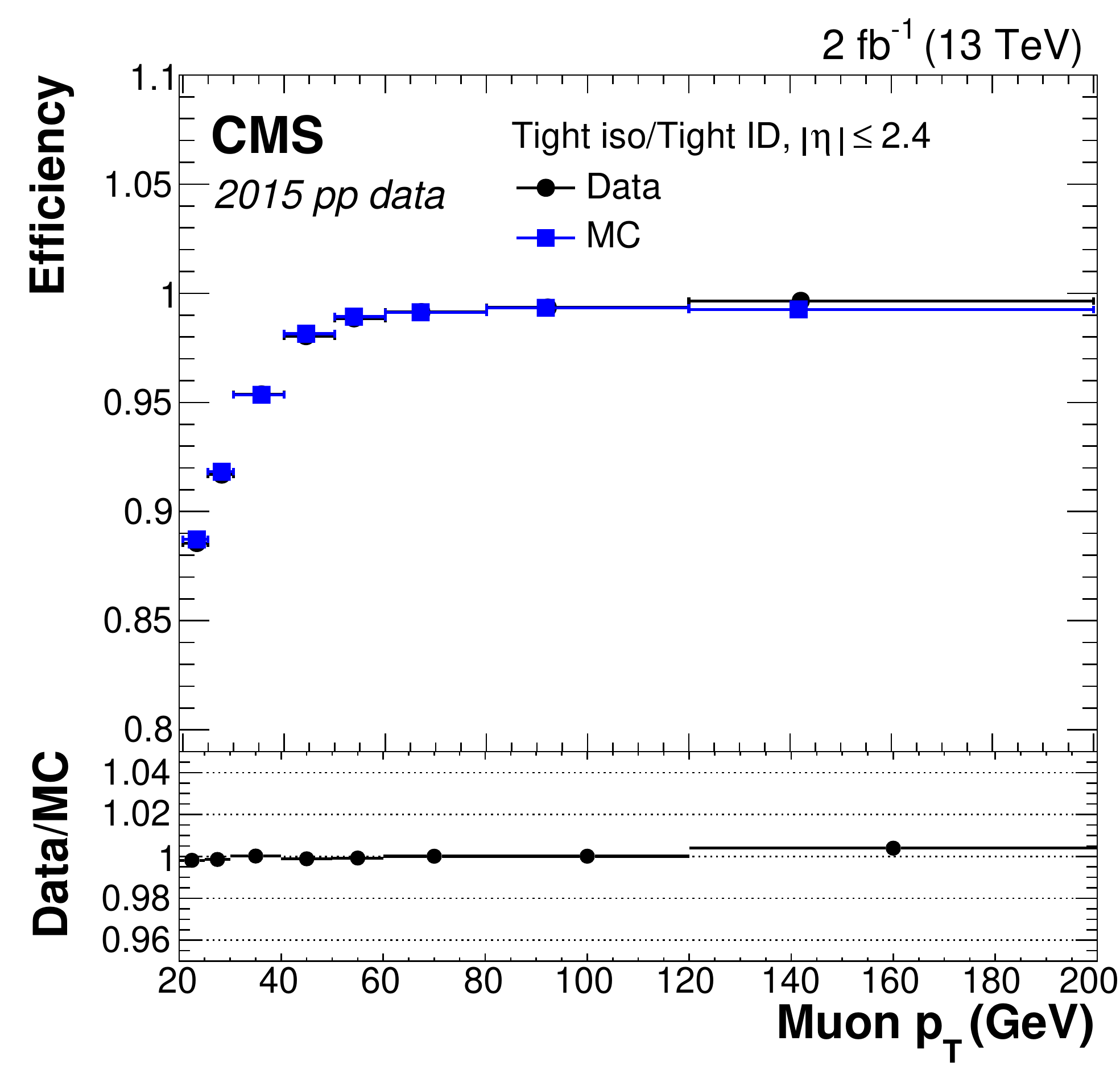} \hspace*{1.0cm}
    \includegraphics[width=0.45\textwidth]{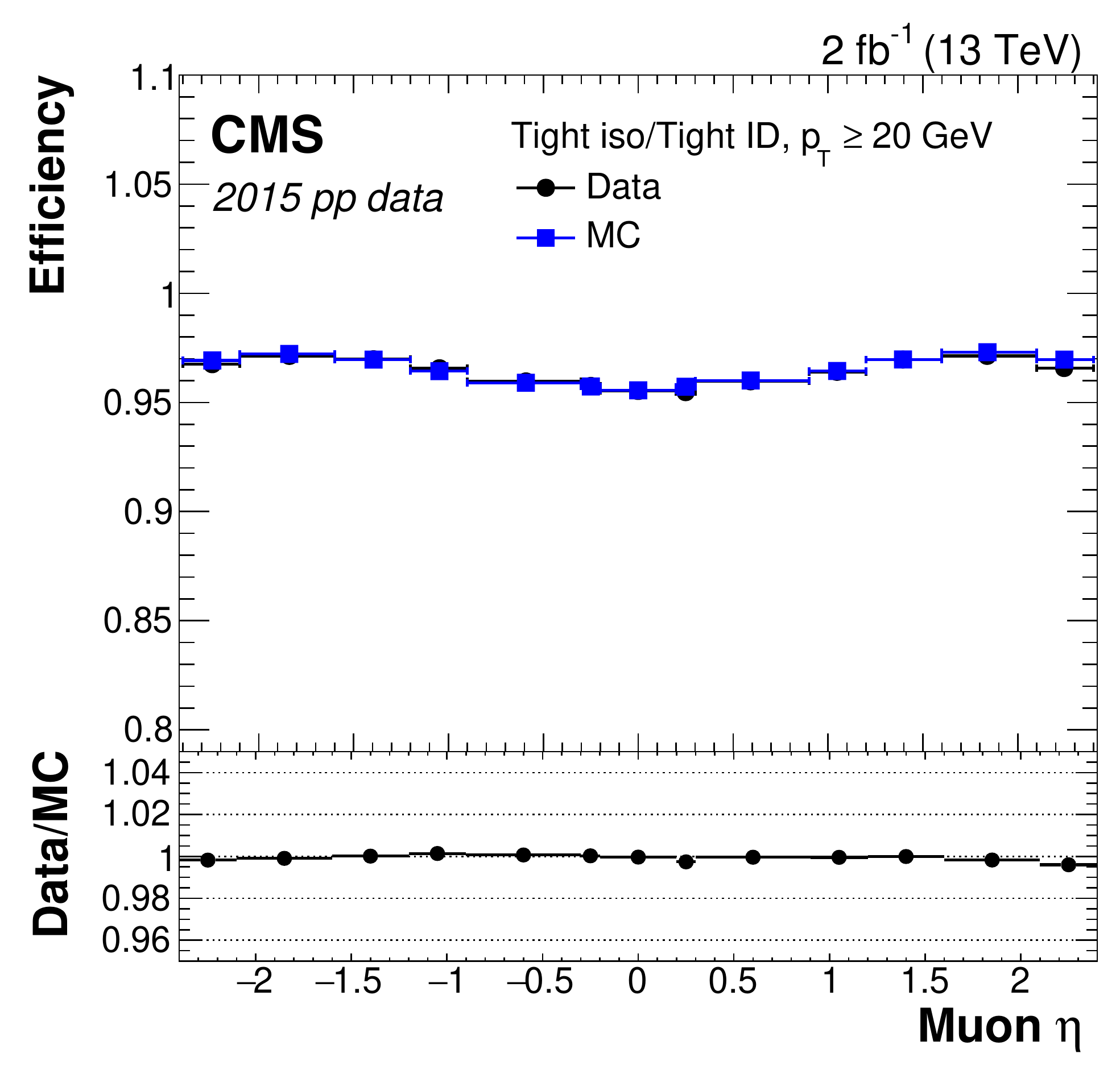}
    \caption{Tag-and-probe efficiency for the tight PF isolation working point on
      top of the tight ID (left) versus \pt for muons in the acceptance of the muon
      spectrometer, and (right) versus pseudorapidity for muons with $\pt > 20\GeV$, for
      2015 data (circles), simulation (squares), and the ratio (bottom inset).
      The statistical uncertainties are smaller than the symbols used to display the
      measurements.
    }
    \label{fig:TightIsoEfficiency}
  }
\end{figure}

The efficiency of muon isolation, $\epsilon_{\text{iso}}$, is studied relative to
a probe that passes a given muon ID criteria.
For example, the tight PF isolation efficiency relative to tight muons is shown
in Fig.~\ref{fig:TightIsoEfficiency}.
In this case the agreement between the data and simulation is always better than
0.5\%.
Analogous to the misidentification probability study described above, the
efficiency to incorrectly label muons within jets as being isolated is measured
with simulated QCD events enriched in muon decays.
In this sample, the probability of a muon with $\pt > 20\GeV$ that fulfills the
tight muon ID criteria to also satisfy tight isolation requirements is about 5\%
in the barrel, and goes up to about 15\% in the endcap.

The systematic uncertainty in data/simulation scale factors for the
efficiencies described above is estimated by varying the tag-and-probe
conditions.
The impact of the background contamination is estimated by using different
requirements on the tag muon (\pt and isolation) and on the requirement of a
single probe being associated with the tag.
The dominant uncertainty is caused by the choice of the signal and background
models used in the fits.
It is estimated by testing alternative fit functions and by varying the range
and the binning of the invariant-mass spectrum.
The uncertainties are estimated to be at the level of 1\% for ID  and 0.5\% for
isolation.

For muons with $\pt > 20\GeV$, Table~\ref{tab:TPplateau} shows the data efficiency
and the data/simulation scale factors for the muon ID and isolation working
points described in Section~\ref{sec:muonreco}.
For all entries, the agreement between data and simulation is better than 1.5\%.
The efficiencies, systematic uncertainties, and scale factors between data and
simulation for 2015 are similar to those found in the 2010 data.
The statistical uncertainties, however, have been reduced by a factor of 10
and become negligible in comparison with the systematic uncertainties.

\begin{table}
  \topcaption{Efficiencies for several reconstruction+ID algorithms and isolation
    criteria (relative to tight ID) for muons with $\pt > 20\GeV$.
    The corresponding scale factors are for 2015 data relative to simulation.
    The uncertainties in the scale factors stem from the statistical uncertainties
    in the fitting procedure.
    Systematic uncertainties are described in the text.
  }
  \centering
  \begin{tabular}{clccc}
    \hline
    Type & Label                     & $\abs{\eta}$ region       & Data eff. [\%]   &  Scale factor        \\
    \hline
    \multirow{8}{*}{Muon ID}
    & \multirow{2}{*}{Loose} & $0.0 < \abs{\eta} < 0.9$ & $99.75 \pm 0.02$ &  $0.998 \pm 0.001$   \\
                             &   & $0.9 < \abs{\eta} < 2.4$ & $99.77 \pm 0.02$ &  $0.9982 \pm 0.0002$ \\
    & \multirow{2}{*}{Medium} & $0.0 < \abs{\eta} < 0.9$ & $98.25\pm 0.02$  &  $0.9901 \pm 0.0002$ \\
                             &   & $0.9 < \abs{\eta} < 2.4$ & $98.55\pm 0.02$  &  $0.9897 \pm 0.0002$ \\
    & \multirow{2}{*}{Tight}    & $0.0 < \abs{\eta} < 0.9$ & $96.00\pm 0.03$  &  $0.9869 \pm 0.0004$ \\
                             &   & $0.9 < \abs{\eta} < 2.4$ & $97.46\pm 0.04$  &  $0.9873 \pm 0.0002$ \\
    & \multirow{2}{*}{High-\pt} & $0.0 < \abs{\eta} < 0.9$ & $96.24\pm 0.02$  &  $0.9882 \pm 0.0003$ \\
    &                           & $0.9 < \abs{\eta} < 2.4$ & $98.05\pm 0.01$  &  $0.9891 \pm 0.0002$ \\[\cmsTabSkip]
    \multirow{4}{*}{Isolation (relative to tight ID)}
    & \multirow{2}{*}{Loose PF} & $0.0 < \abs{\eta} < 0.9$ & $98.60 \pm 0.01$ &  $1.0007 \pm 0.0001$ \\
    &                           & $0.9 < \abs{\eta} < 2.4$ & $98.98 \pm 0.01$ &  $1.0007 \pm 0.0001$ \\
    & \multirow{2}{*}{Tight PF} & $0.0 < \abs{\eta} < 0.9$ & $95.81 \pm 0.02$ &  $1.0001 \pm 0.0004$ \\
    &                           & $0.9 < \abs{\eta} < 2.4$ & $96.88 \pm 0.02$ &  $0.9995 \pm 0.0003$ \\
    \hline
  \end{tabular}
  \label{tab:TPplateau}
\end{table}

\section{Momentum scale and resolution}
\label{sec:momentumscale}

Many searches for new physics are characterized by signatures involving prompt muons
with high \pt.
For muons with $\pt > 200\GeV$, combining information from the muon system with
information from the inner tracker significantly improves the momentum
measurement~\cite{PhysicsTDR}.
On the other hand, for muons with lower \pt the momentum measurement is dominated by
the performance of the inner tracker.
To assess the performance of the momentum scale and resolution, data from both cosmic
rays and collisions have been analyzed.

\subsection{Low and intermediate \texorpdfstring{\pt}{pT}:  scale and resolution with collisions}

For muons with low and intermediate \pt, two different methods are utilized in Run~2
to correct the muon momentum scale and to estimate the resolution.
One method derives the corrections from the mean value of the distribution of
$1/\pt^{\mu}$, $\left < 1/\pt^{\mu} \right >$, for tight muons from {\cPZ} decays, with
further tuning performed using the mean of the dimuon invariant mass spectrum,
$\left < M_{\mu\mu} \right >$~\cite{Rochester}.
Another method determines corrections using a Kalman filter on tight muons from
\PJGy and \PgUa~decays~\cite{Kalman}.
The magnitudes of the momentum scale corrections are about 0.2\% and 0.3\% in the
barrel and endcap, respectively.
After the scale is corrected, the resolution is determined either as a function of
$\eta$ (first method) or as a function of $\eta$ and \pt (second method), including
contributions from multiple scattering, position error, and additional smearing to
make the simulation match the data.
The resolution for muons with momenta up to approximately 100\GeV is 1\% in the
barrel and 3\% in the endcap.
For both techniques, over all $\eta$ and \pt values, the uncertainty in the
resolution is estimated to be about 5\% of its value.
Compared to the 2010 results~\cite{POGPerformance}, the 2015 resolution has
improved, primarily because of the improvements to the tracker
alignment~\cite{TrackerAlignment}.

\subsection{Momentum resolution with cosmic rays}

Cosmic ray muons passing through the CMS detector are used to estimate the momentum
resolution at high \pt by comparing the momentum measured in the upper half of the
detector with the momentum measured in the lower half~\cite{POGPerformance,CRAFT}.
Events are selected with muons that cross the detector close to the interaction
point and have at least one hit in the pixel detector, so that each leg of the
cosmic ray mimics a muon from a collision.
To ensure good reconstruction, the tracker track of each muon leg is required to have
at least one pixel hit as well as five strip layers.
The relative \textit{q}/\pt residual, $R(q/\pt)$, is computed as
\begin{equation}
  R(q/\pt) = \frac{1}{\sqrt{2}}\frac{(q/\pt)_{\text{upper}} - (q/\pt)_{\text{lower}}}{(q/\pt)_{\text{lower}}},
\end{equation}
where \textit{q} is the muon charge, and upper and lower refer to the muon tracks
reconstructed in the upper and lower halves of the CMS detector, respectively.
The quantity \textit{q}/\pt, proportional to the muon trajectory curvature, has a
symmetric, approximately Gaussian, resolution distribution.
The factor of $\sqrt{2}$ accounts for the fact that the \textit{q}/\pt measurements
of the two tracks are independent.

\begin{figure}[ht]
  \centering
  \includegraphics[width=0.8\linewidth]{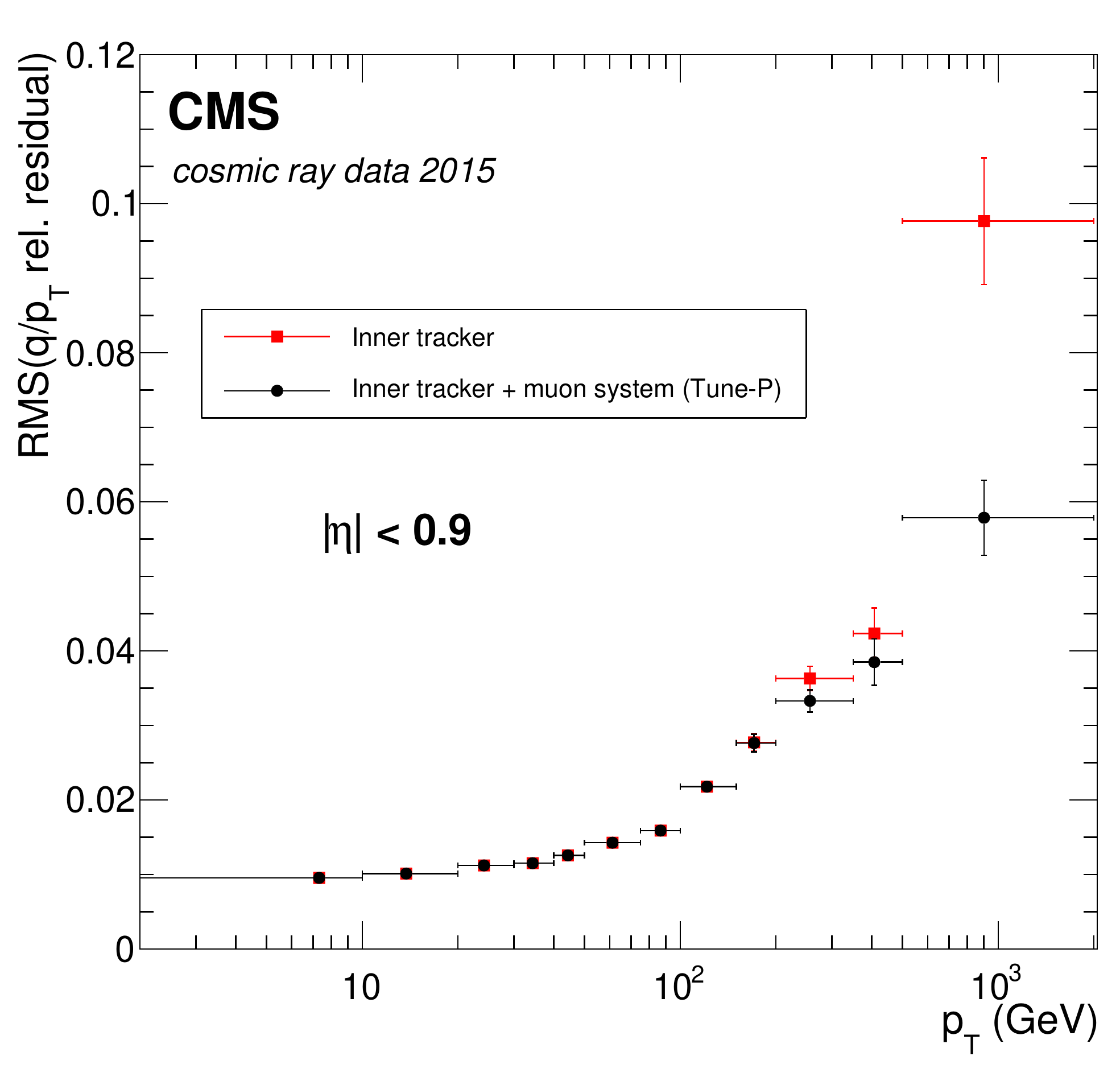}
  \caption{The RMS of $R(q/\pt)$ as a function of \pt for cosmic rays recorded in
    2015, using the inner tracker fit only (squares) and including the muon system
    using the \textit{Tune-P} algorithm (circles).
    The vertical error bars represent the statistical uncertainties of the RMS.
  }
  \label{fig:ResolutionvsPt}
\end{figure}

Figure~\ref{fig:ResolutionvsPt} shows the RMS of $R(q/\pt)$ as a function of \pt
for cosmic rays recorded in 2015 for fits using only the inner tracker and for fits
that include the muon system using the \textit{Tune-P} algorithm.
The uncertainty in the last bins is dominated by the small number of cosmic rays
collected in 2015 (66 events with $\pt > 500\GeV$).
The improvement in resolution from exploiting the muon chamber information in the
momentum assigment is clearly visible.
The simulation of cosmic rays with $\pt > 500\GeV$ reproduces this result within
statistical uncertainties.
Compared with the 2010 results, the resolution is improved by about 25\% at high
\pt, coming as a result of the modifications to the \textit{Tune-P} algorithm in
addition to the improved alignment of both the inner
tracker~\cite{TrackerAlignment} and the muon system~\cite{DPGPerformance}.

\subsection{High \texorpdfstring{\pt}{pT}:  momentum scale with collisions}

Biases in the scale of the momentum measurement at high \pt arising from an
inaccurate measurement of the track curvature are probed by looking for
distortions in the shape of the \textit{q}/\pt spectrum.
A technique called the ``endpoint method'' was developed and used extensively in
Run~1, using cosmic ray data to quantify the bias at high
\pt~\cite{POGPerformance,DPGPerformance}.
However, since cosmic rays predominantly cross the barrel region of the detector,
they cannot be used effectively to determine the momentum scale in the endcaps.
Therefore, a generalized version of the endpoint method has been developed to be
used with collisions.

The generalized endpoint method uses prompt dimuons selected from a sample of
events collected with the single-muon trigger (see Section~\ref{sec:trigger}).
Both muons must satisfy the loose tracker relative isolation criteria and at
least one of the muons is required to have $\pt > 200\GeV$.
This sample is primarily composed of muons from \cPZ/\PGg$^*$ decays,
with a minor contribution from dileptonic decays of \ttbar pairs and from
diboson production.

Each muon from the event that has $\pt > 200\GeV$ is used to fill a binned
distribution of $q$/\pt.
The \textit{q}/\pt data spectrum is compared to multiple samples of simulated
muons.
Each sample, i, is simulated with a curvature bias, $k^{\mathrm{i}}_{\mathrm{b}}$,
injected on top of an unbiased geometry.
The bias shifts the unbiased \textit{q}/\pt spectrum by
\begin{equation}
  q/\pt \to q/\pt + k^{\mathrm{i}}_{\mathrm{b}}.
\end{equation}
The samples are generated with $k^{\mathrm{i}}_{\mathrm{b}}$ in steps of 0.01/\TeV
between $-$1.00/\TeV and $+$1.00/\TeV.
For each sample, the $\chi^2$ is computed between the unweighted data
distribution and the weighted simulation distribution.
The value of $\chi^2$ is plotted as function of $k^{\mathrm{i}}_{\mathrm{b}}$ and fit
with a second-degree polynomial.
The value of $k^{\mathrm{i}}_{\mathrm{b}}$ that gives the minimum fit $\chi^2$ is
taken as the curvature bias in the data, $k_{\mathrm{b}}$.
The statistical uncertainty in $k_{\mathrm{b}}$ corresponds to half the range
over which the $\chi^2$ increases by one.

The momentum scale bias in 2015 data from the generalized endpoint method is
shown in Table~\ref{tab:MomentumScale}.
The bias is presented separately for the barrel and endcaps and integrated over
$\phi$.
Within the statistical uncertainties, the measurements are consistent with no
bias.
In both the barrel and endcaps, the amplitude of the azimuthal dependence of
$k_{\mathrm{b}}$ is less than 0.1/\TeV.
The limited statistical precision of the data precludes detailed studies of
the $\phi$ dependence and a detailed analysis of the width of $k_{\mathrm{b}}$.
An analysis using the cosmic ray endpoint method in the barrel is consistent
with Tab.~\ref{tab:MomentumScale}.
However, the large uncertainties in the cosmic ray data don't constrain the bias
better than the collision data alone.
The scale bias in the 2015 data is approximately consistent with the scale bias
measured in 2010 with cosmic rays~\cite{POGPerformance}, within the large
uncertainties in the 2010 results.

\begin{table}
  \topcaption{Measurement of the momentum scale bias in 2015 data, obtained with
    the generalized endpoint method using muons with $\pt > 200\GeV$ from
    {\Pp\Pp} collision data.
    Results are presented in three $\eta$ bins corresponding to the barrel and
    endcap regions.
  }
  \centering
  \begin{tabular}{lccc}
    \hline
    $\eta$ range                           & $-2.4 < \eta < -1.2$ & $-1.2 < \eta < 1.2$ & $1.2 < \eta < 2.4$  \\
    $\langle k_{\mathrm{b}} \rangle$ (1/\TeVns{}) & $-0.01 \pm 0.06$     & $-0.01 \pm 0.03$    & $0.01 \pm 0.05$ \\
    \hline
  \end{tabular}
  \label{tab:MomentumScale}
\end{table}

\section{Timing}
\label{sec:timing}

The ``L1 accept'' signal, which is broadcast to all subdetectors, initiates the readout of
the event.
Trigger synchronization is of great importance because as simultaneous hits in multiple
chambers are required for an L1 trigger, out-of-time chambers can reduce the overall
trigger efficiency.
Moreover, if the L1 muon trigger is generated early or late relative to the collision time,
it forces readout of the entire detector at the wrong bunch crossing.
In this context, the timing performance of the RPC hits and the DT and CSC trigger
primitives is discussed in Section~\ref{subsec:triggerprimitives}.

For physics analyses, the time assigned to the muon hits once the event has been collected
and fully reconstructed is also important.
This is called the ``offline time.''
For a muon traveling at the speed of light, produced in a proton-proton collision, and
with the correct bunch crossing assignment, the offline time of any muon chamber hit
should be reported as $t=0$.
The readout windows of the muon subsystems are large enough to detect muons from several
bunch crossings.
Any deviations from 0 may be caused by backgrounds such as cosmic-ray muons, beam
backgrounds, chamber noise, or out-of-time pileup, or it may be an indication of new
physics such as a slow moving, heavy charged particle.

As described in Section~\ref{sec:muonreco}, the timing information of DT segments is
obtained from a 3-parameter fit of segments, so that position, direction and time of a
crossing track are determined simultaneously.
Single track segments were selected to have hits in both projections (at least five in
the $\phi$ view) and to have an inclination angle below 45\de.
The $\sigma$ parameter of a Gaussian fit to the segment time distribution is 2.0\unit{ns},
which represents an estimate of the DT segment time resolution.
An improvement of about 0.6\unit{ns} is observed with respect to the 2010
performance~\cite{DPGPerformance}.
This improvement results from the updated segment reconstruction algorithm that now
explicitly measures the segment time.

\begin{figure}[ht]
  \centering
  \includegraphics[width=0.80\linewidth]{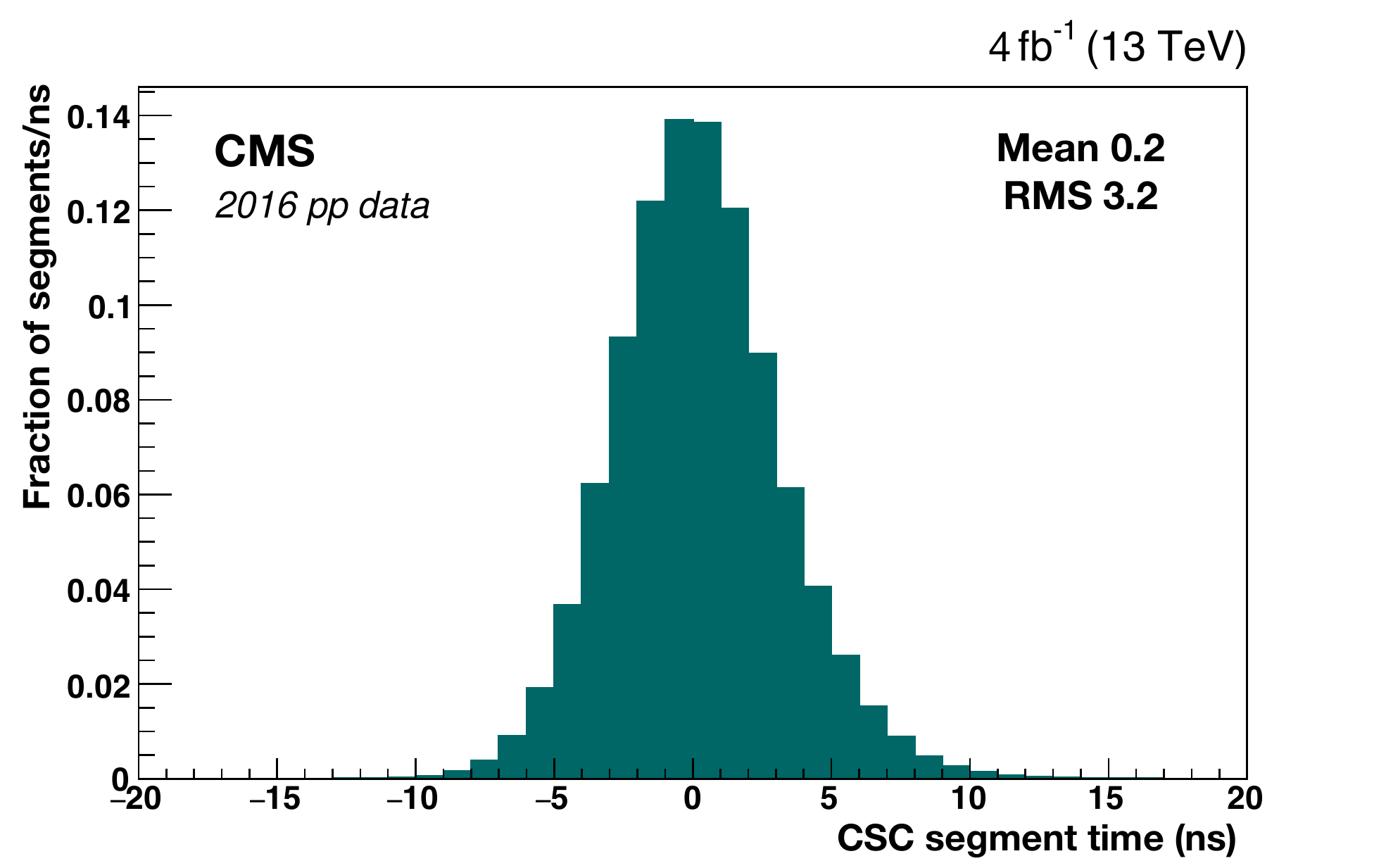}
  \caption{Distribution of times from reconstructed CSC segments measured with the
    2016 data.
  }
  \label{fig:dpg6}
\end{figure}

The time of a CSC reconstructed segment is determined by combining the times of the
cathode and anode hits used to construct the segment.
The overall precision depends mostly on the cathode timing performance.
The cathode time is determined from a template fit to the digitized cathode pulse.
It is calibrated based on dedicated studies of chamber response and a heuristic
correction measured from collision data.
Figure~\ref{fig:dpg6} shows the distribution of times of CSC segments associated with
reconstructed muons.
The RMS of the binned segment time distribution is 3.2\unit{ns}, in good agreement
with the value of about 3\unit{ns} measured in 2010~\cite{DPGPerformance}.

\begin{figure}[ht]
  \centering
  \includegraphics[width=0.80\linewidth]{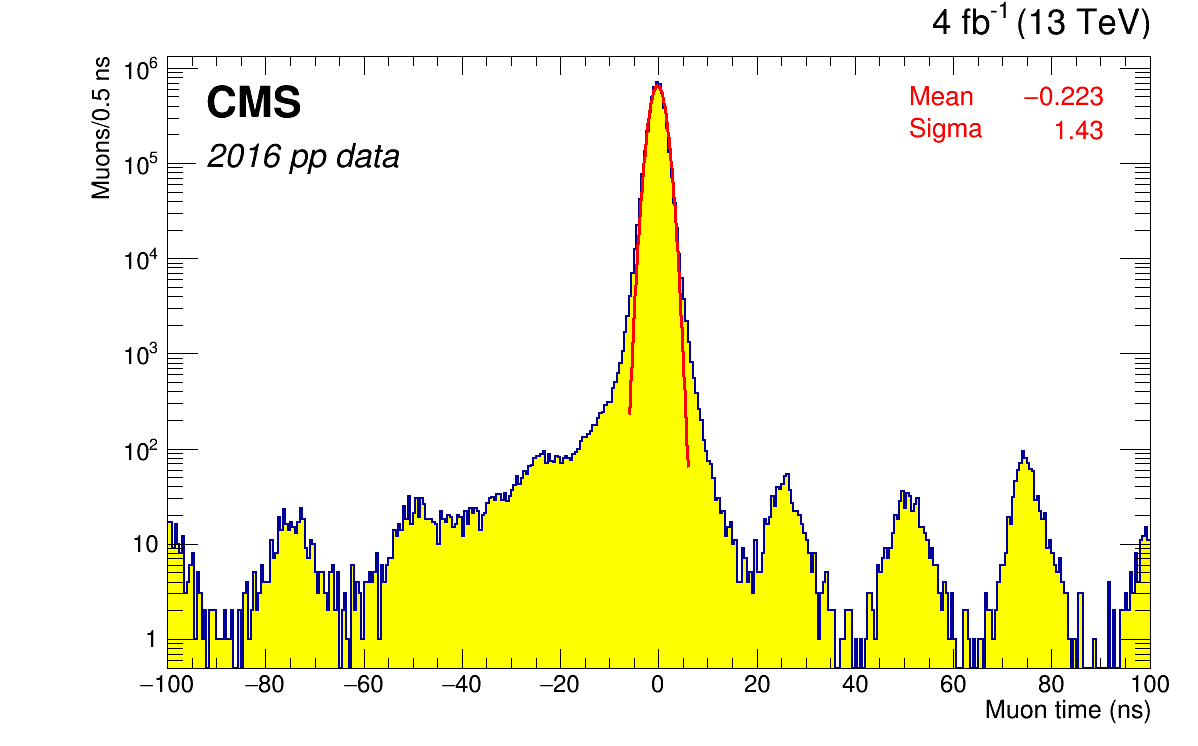}
  \caption{Time-at-vertex distribution for standalone-muons in the barrel, using
    the times measured by DT chambers in 2016 data.
  }
  \label{fig:dpg7}
\end{figure}

The timing of a standalone-muon can be determined by combining measurements from
multiple stations.
In the barrel, measurements from up to four DT stations are combined using an
iterative pruning mechanism to discard outlier hits from those associated with the
track, thereby rejecting hits from delta rays and showers within an individual
chamber.
The time-at-vertex distribution for standalone-muons in the barrel is shown in
Fig.~\ref{fig:dpg7}.
This distribution comes from muons that have triggered the event readout and shows a
primary peak clearly visible at 0\unit{ns}.
The asymmetric tail at early values comes from delta rays that reach the wire before
the hit.
The peaks periodically spaced at 25\unit{ns} both before and after the primary peak
come from muons produced in LHC collisions in bunch crossings that are out of time
with regard to the trigger.
These secondary peaks come from muons that did not trigger the event readout because
of the large suppression factors in the CMS trigger system (described in
Section~\ref{sec:trigger}) but are within the readout range of the event that did
cause the trigger.
A comparison with an analysis from Run~1 data~\cite{HSCPPaper} shows that in 2016
data the width of the primary timing peak in the muon barrel was consistent within
0.5\unit{ns}.
A similar analysis using the CSC shows comparable muon time resolution.

\section{Trigger}
\label{sec:trigger}

For the muon component of the CMS trigger~\cite{L1Paper}, CSC and DT chambers
provide ``trigger primitives'' constructed from hit patterns consistent with
muons that originate from the collision region, and RPC chambers provide hit
information.
The custom-made electronics in the L1 trigger system utilize the chamber
information to reconstruct muon trigger candidates with a coarse measurement of
\pt.
An upgrade of the L1 trigger system was implemented between 2015 and
2016~\cite{CMSPhase1TriggerTDR}.
The results presented here do not depend on the upgraded L1 components.

Events selected by the L1 trigger are passed to the HLT, which uses information
from the full CMS detector to reconstruct muons.
The HLT algorithms are simplified versions of those described in
Section~\ref{sec:muonreco} in order to reduce computing time and resources, and
were updated prior to the 2015 run to improve the performance at higher
pileup.
The HLT employs two different processing schemes to reconstruct muons.
The first scheme uses the L1 candidate as a seed to reconstruct muons at level-2
(L2) using information only from the muon system, and then reconstructs level-3
(L3) muons by combining the L2 muons with information from the inner tracker.
This combination is made using three algorithms applied sequentially from
fastest to slowest, and subsequent algorithms are attempted only if the previous
one failed to reconstruct a muon in order to minimize computation time.
The first algorithm propagates the L2 trajectory inward to the inner tracker to
reconstruct the L3 muon.
The second algorithm is similar to the first, except that it combines the L2
muon with hits in the outer layer of the inner tracker to improve its trajectory
before propagating it inward.
The third algorithm is different from the first two in that it builds tracker
tracks with an inside out approach within a region based on the position of the
L2 muon.
Prior to Run~2, the L3 algorithms were improved to select hits based on $\chi^2$
of the track fit rather than matching in $\Delta R$, and track quality
constraints are imposed in the first two algorithms.

The second HLT processing scheme, called ``HLT tracker muon reconstruction'',
was developed prior to Run~2.
This scheme employs an algorithm similar to the tracker muon algorithm described
in Section~\ref{sec:muonreco}, but is optimized for processing speed.
The primary differences are that the HLT version limits the reconstruction of
tracker tracks to a region within $\Delta \phi < 0.2$ and $\Delta \eta < 0.35$
of the L1 candidate, and requires $\pt > 10\GeV$ for the tracker track seeds.

After reconstruction, muon isolation is evaluated in the HLT by considering the
additional tracker tracks and calorimeter energy deposits in a cone with radius
$\Delta R = 0.3$ around the muon.
Each of the contributions is required to be below a fraction of the muon \pt:
scalar sum \et of PF electromagnetic clusters~\cite{PRF-14-001}, scalar sum \et
of PF hadronic clusters, and scalar sum \pt of tracker tracks.
To exclude contributions that come from the muon itself, a minimum value of
$\Delta R$ is required to include the tracks or energy deposits in the sum.
To account for the effects from pileup, PF cluster sums are corrected using the
average energy density~\cite{FastJet} in the event, $\rho$ (if the correction
exceeds the PF cluster sum, that component of the isolation is set to zero).
To determine the correction, the value of $\rho$ is scaled by its
``effective area'' which estimates how much is expected in the isolation cone.
Effective areas are determined independently for electromagnetic sums and for
hadronic sums as well as separately in the barrel and in the endcaps.
The average values of the distributions of PF cluster sums are fit with a first
order polynomial as a function of the number of reconstructed primary vertices.
The same is done for $\rho$.
The effective area is the ratio of the fitted slope for the PF cluster sum
divided by the fitted slope for $\rho$.

After minimal $\Delta R$ cones and effective areas are defined, a working point
is determined to simultaneously remove background effectively and to keep signal
efficiency high by tuning the thresholds below which the muon is considered to
be isolated.
For example, for online isolation in the barrel in 2015, the $\rho$-corrected
scalar sum \et of PF electromagnetic clusters within $0.05 < \Delta R < 0.3$
were required to be below 11\% of the muon \pt, the $\rho$-corrected scalar sum
\et of PF hadronic clusters within $0.1 < \Delta R < 0.3$ were required to be
below 21\% of the muon \pt, and the scalar sum \pt of tracker tracks within
$0.01 < \Delta R < 0.3$ were required to be below 9\% of the muon \pt.

The results of the HLT reconstruction and isolation algorithms are used to form
various trigger conditions.
The general-purpose muon trigger conditions used for the 2015 data include:
\begin{enumerate}
\item an isolated single-muon with a \pt threshold of 20\GeV, which is based on a
  trigger efficiency curve giving approximately 50\% efficiency at
  20\GeV~\cite{L1Paper}, reconstructed with either L3 or HLT tracker muon
  algorithms,
\item a nonisolated single-muon with a \pt threshold of 45\GeV for $|\eta| < 2.1$
  or 50\GeV for $\abs{\eta} < 2.4$, reconstructed with L3, and
\item two isolated muons (double-muons) that originate within a distance of
  $\Delta z < 0.2$\cm of each other along the beamline, with asymmetric \pt
  thresholds of 18\GeV and 7\GeV applied to the two muons.
\end{enumerate}

For the double-muon triggers, the L3 algorithm is first used to reconstruct one
muon.
In order to save computing time, this L3 muon must pass \pt and quality
constraints before reconstruction of a second muon is attempted.
The second muon can be reconstructed with either the L3 or the HLT tracker
muon algorithm to maximize efficiency.
Tracker track isolation criteria are then applied to both tracks.

\subsection{Trigger primitives}
\label{subsec:triggerprimitives}

\begin{figure}[ht]
  \centering
  \includegraphics[width=0.80\linewidth]{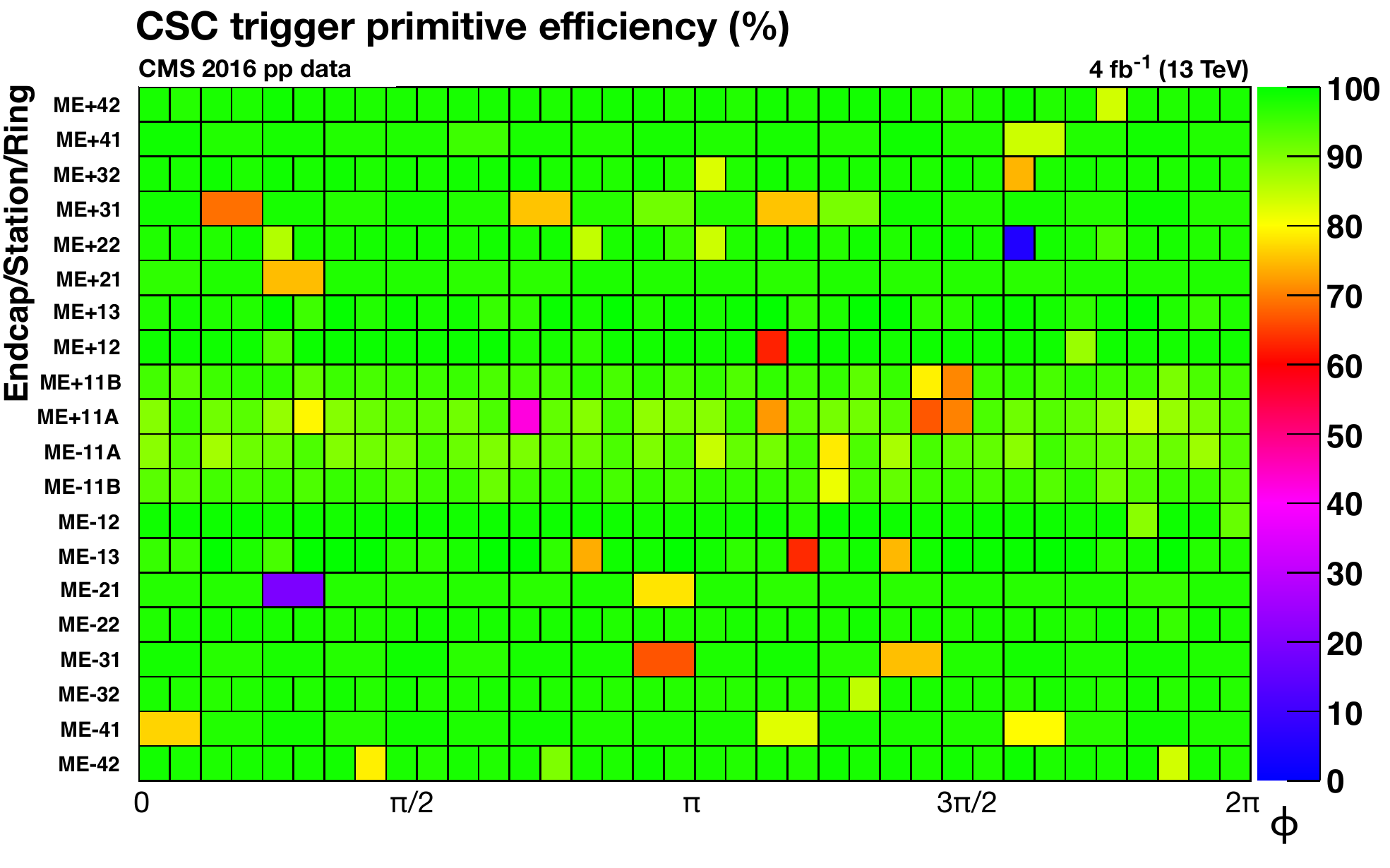}
  \caption{The efficiency (in percent) of each CSC to provide a trigger primitive,
    measured with the 2016 data.
  }
  \label{fig:dpg9}
\end{figure}

The absolute efficiency for creating a CSC trigger primitive is studied using
the tag-and-probe method in the same way as for CSC segments
(Section~\ref{sec:efficiency}).
Once again the probe is a tracker track extrapolated into the muon system.
The tag is required to have triggered the event to avoid bias from events
triggered by the probe alone.
A CSC trigger primitive is expected in each chamber traversed by the probe.
To reduce background and to ensure that the probe actually entered the chamber
under consideration, a compatible segment is required in a downstream chamber.
For the outermost station 4 an upstream chamber is required instead.
A trigger primitive is required to be within 5\cm of the extrapolated track
(corresponding to about 4--5 times the resolution, as demonstrated in
Fig.~\ref{fig:MuonResiduals}b) with no other track closer to it.
The CSC trigger primitive efficiency is shown in Fig.~\ref{fig:dpg9}.
The features in Fig.~\ref{fig:dpg9} are highly correlated with the features
in Fig.~\ref{fig:dpg5} because in both cases the primary causes of significant
inefficiencies were hardware failures.
The average CSC trigger primitive efficiency in 2016 data is 97\%, similar to
that in 2010~\cite{DPGPerformance}.

\begin{figure}[ht]
  \centering
  \includegraphics[width=0.80\linewidth]{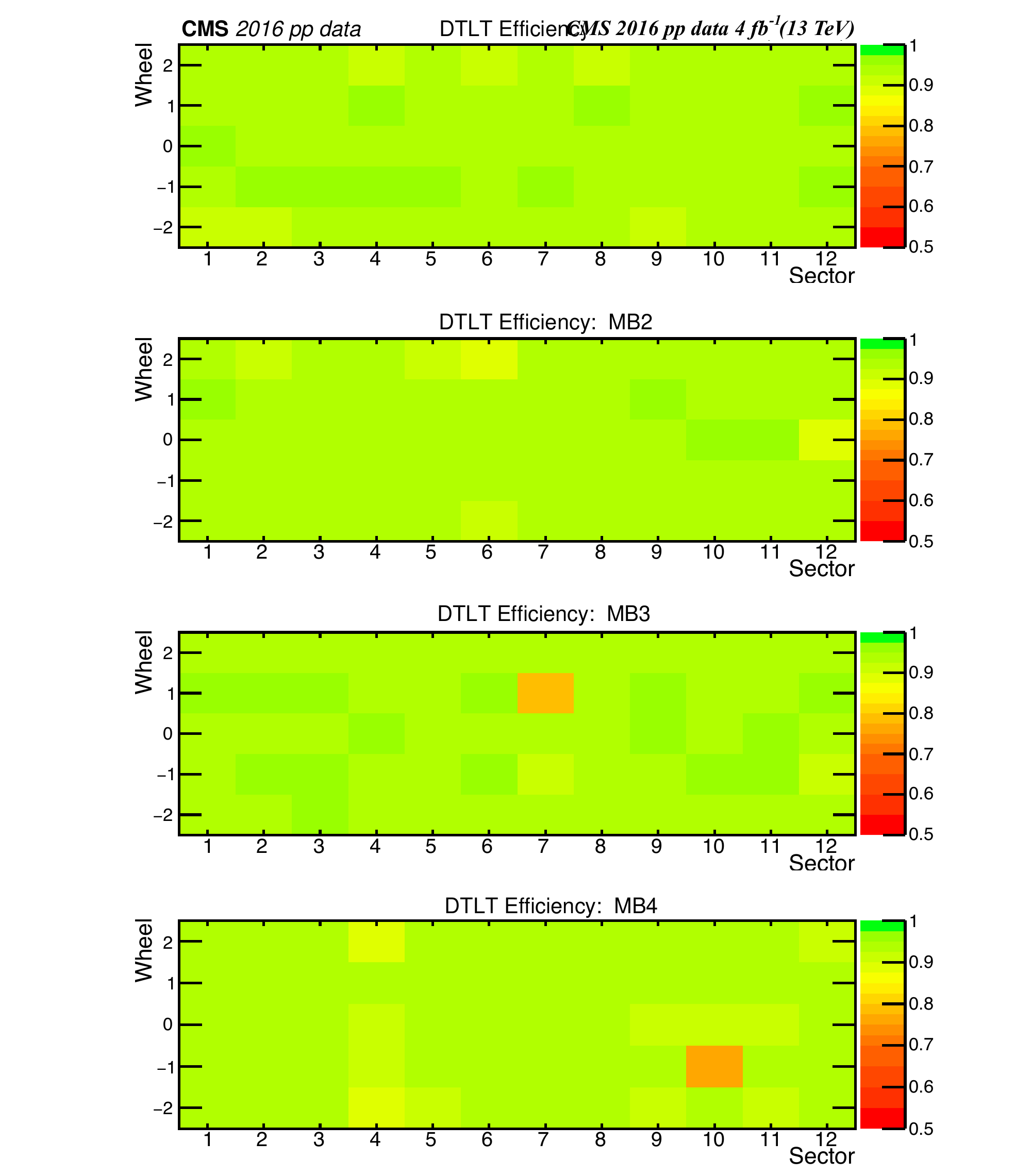}
  \caption{Efficiency map for the DT local trigger ($\phi$ view) for each chamber,
    measured with the 2016 data.
    Each map represents one station.
    The z-axis color indicates the efficiency, the wheel number is on the vertical
    axis, and the $\phi$ sector number is on the horizontal axis.
  }
  \label{fig:dpg8}
\end{figure}

The efficiency for the DT local trigger electronics to reconstruct a trigger
primitive pattern is called the DTLT efficiency, and it can be studied using
segments associated with global muons.
In order to ensure that the chamber under study was not necessarily used to
trigger the event, at least two other stations are required to deliver trigger
primitives.
The denominator is incremented if the segment is reconstructed in both $\theta$
and $\phi$ views, except for MB4, which has only $\phi$ superlayers.
In addition, there must be at least four associated hits in the $\phi$ layers,
the minimum number of hits required to build a $\phi$ trigger primitive.
The numerator is incremented if a trigger primitive is delivered at the correct
bunch crossing.
The DTLT efficiency is shown for each DT chamber in Fig.~\ref{fig:dpg8}.
The lower DTLT efficiency observed in two of the chambers was due to problems
with the trigger electronics which were later repaired.
The DTLT efficiency is about 1\% lower in MB4 because there are no $\theta$
superlayers to enhance the quality of the segment.
The DTLT efficiency in the 2016 data is comparable to the one observed in the
2010 data.

For RPCs, the trigger primitive efficiency is equivalent to the hit efficiency by
construction.  The trigger primitive efficiency is shown in Fig.~\ref{fig:dpg1}.

\begin{figure}[ht]
  \centering
  \includegraphics[width=0.45\linewidth]{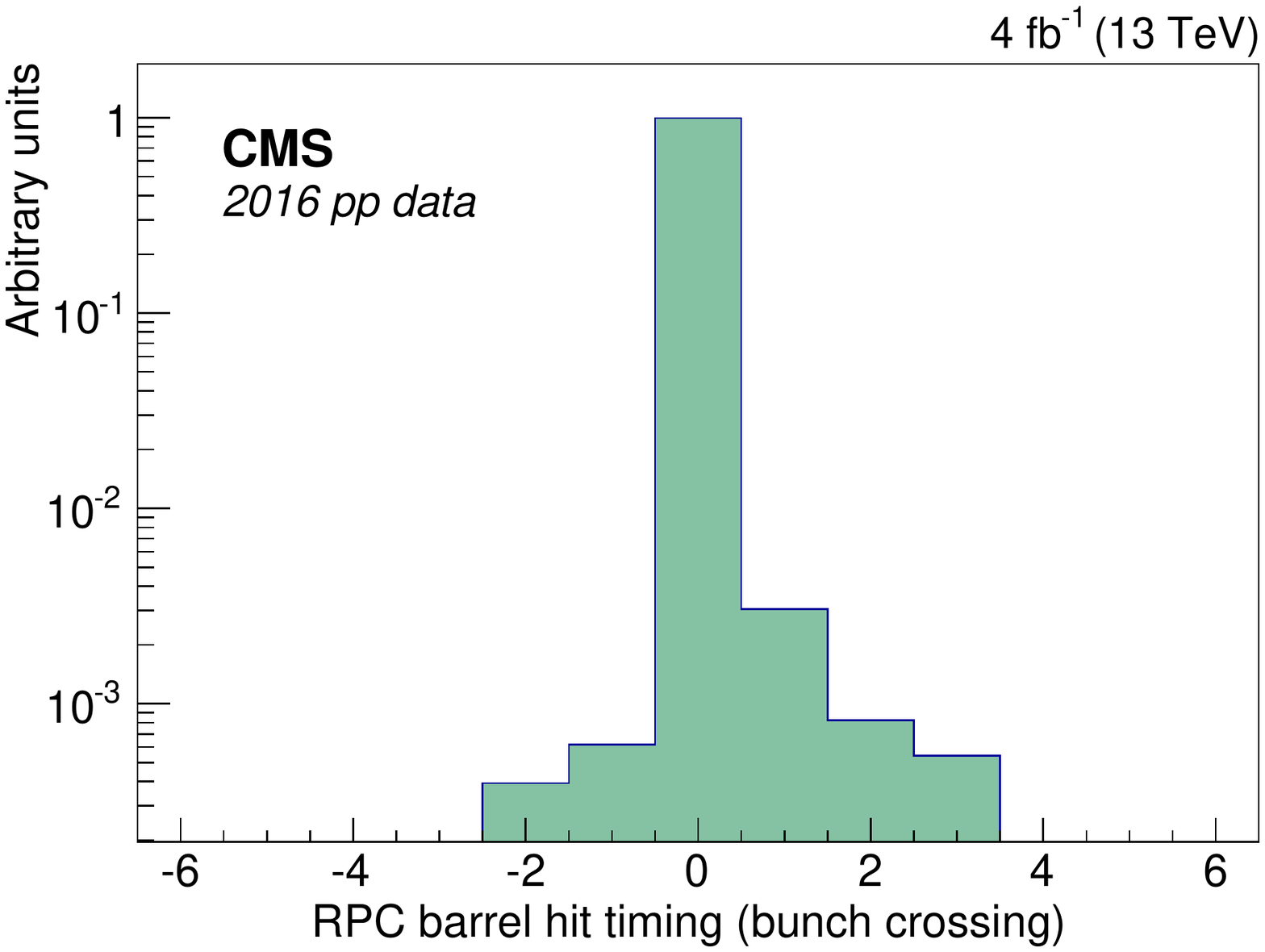} \hfill
 \includegraphics[width=0.45\linewidth]{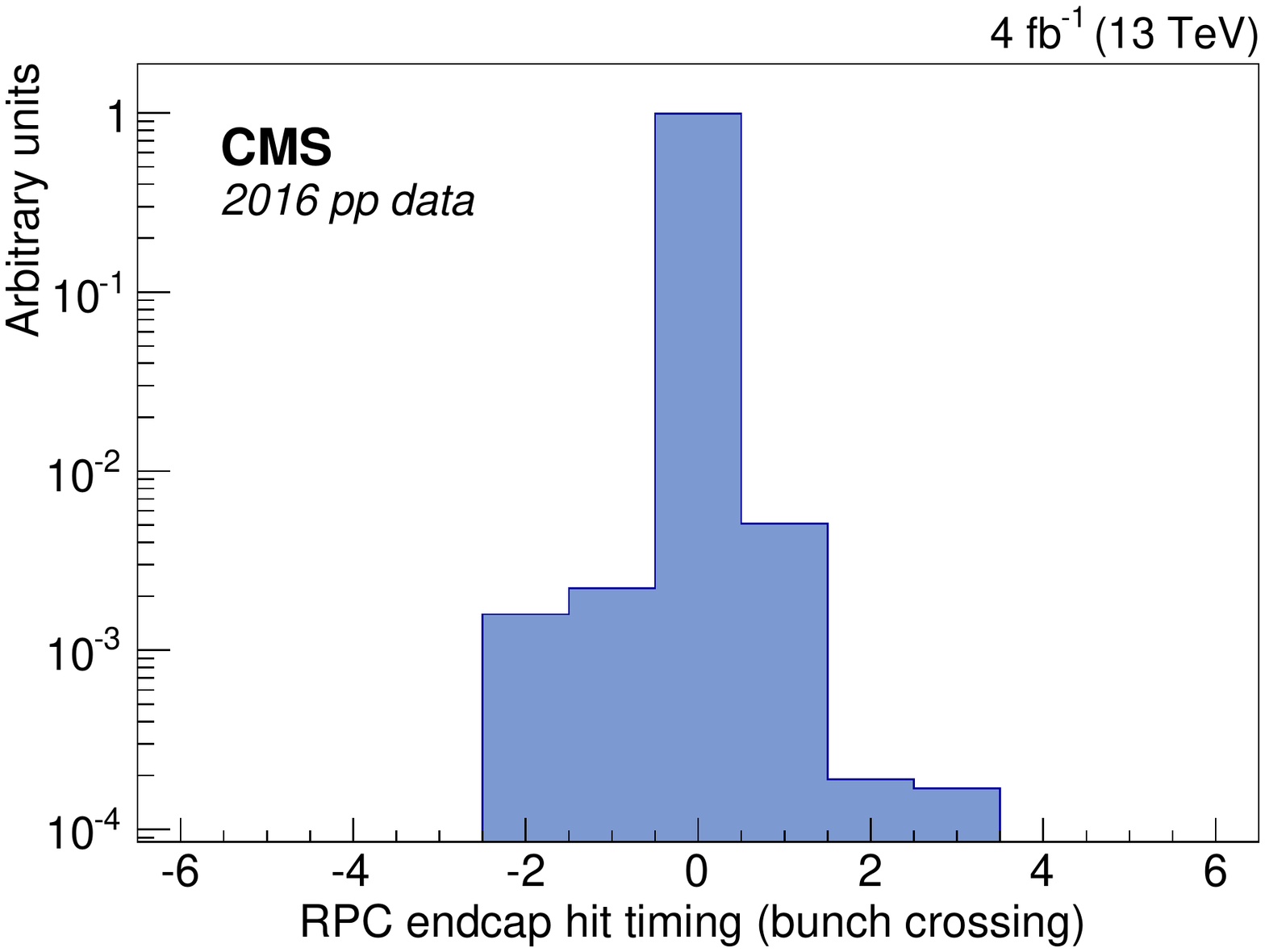} \caption{The bunch crossing distribution from reconstructed RPC hits in the barrel
    (left) and in one endcap (right), using the 2016 data.
  }
  \label{fig:dpg10}
\end{figure}

The time coincidence of trigger primitives among the many muon stations must have
a time dispersion much less than 25\unit{ns}, the time separation of LHC bunch
crossings, to ensure an unambiguous identification of the correct bunch crossing
with the muon trigger.
For example, the RPC chambers have been measured to have an intrinsic time
resolution of around 2\unit{ns}~\cite{RPCtiming} and an overall time resolution
of better than 3\unit{ns}~\cite{DPGPerformance} after including the time
propagation along the strip, the channel-by-channel cable length differences, and
the electronics delays.
Figure~\ref{fig:dpg10} shows the bunch crossing distribution of RPC hits
associated with global muons in the barrel.
Each bin corresponds to the 25\unit{ns} bunch separation in LHC, and bin 0 is the
time of the L1 trigger.
In Fig.~\ref{fig:dpg10}, 0.5\% of RPC hits are outside bin~0, whereas for both DT
and CSC trigger primitives, 2\% are outside bin~0.
The hits that are not in bin~0 are caused by a combination of muons from adjacent
bunch crossings or from cosmic rays and by the finite resolution in the
calibration of the electronics.
The timing of each individual system is monitored during data collection and fine
adjustments are made if necessary.
In this way, the L1 trigger, which relies on a combination of all three systems,
produces less than 0.2\% trigger candidates associated with incorrect bunch
crossings~\cite{L1Paper}.

\subsection{Trigger efficiencies and rates}
\label{sec:muontriggerreco}

\begin{figure}[ht]
  \centering
 \includegraphics[width=0.45\textwidth]{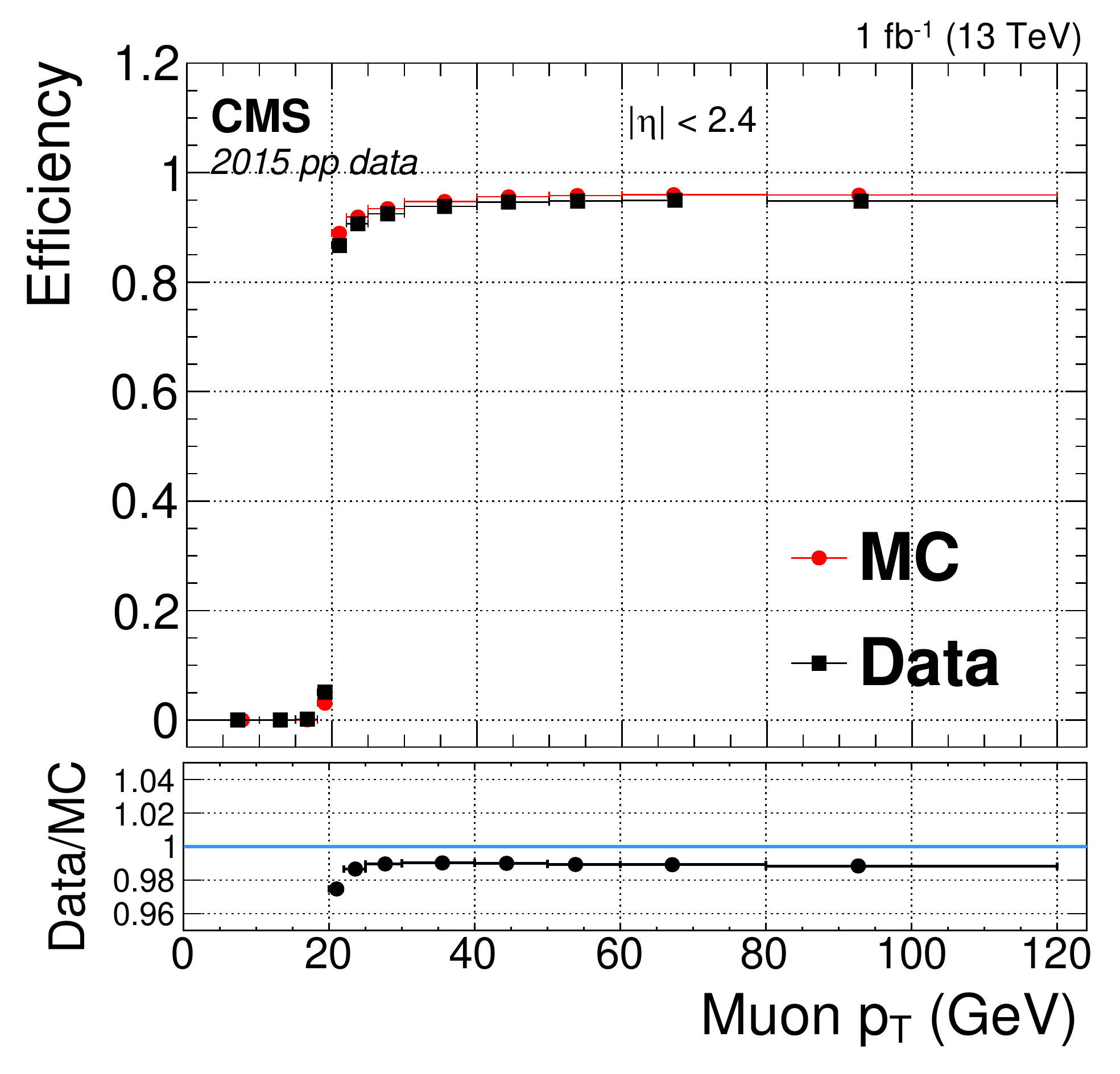}
 \includegraphics[width=0.45\textwidth]{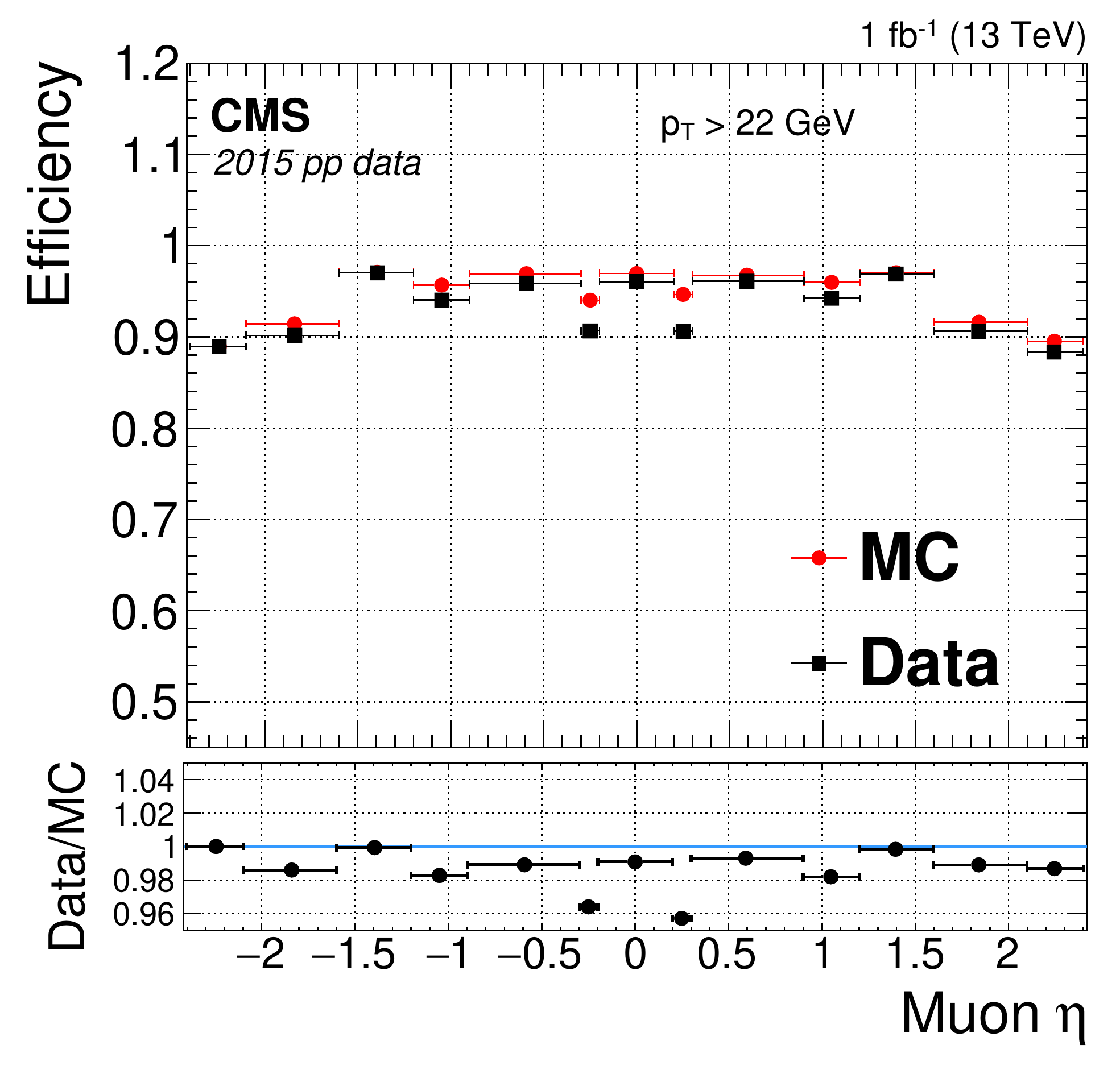}

 \includegraphics[width=0.45\textwidth]{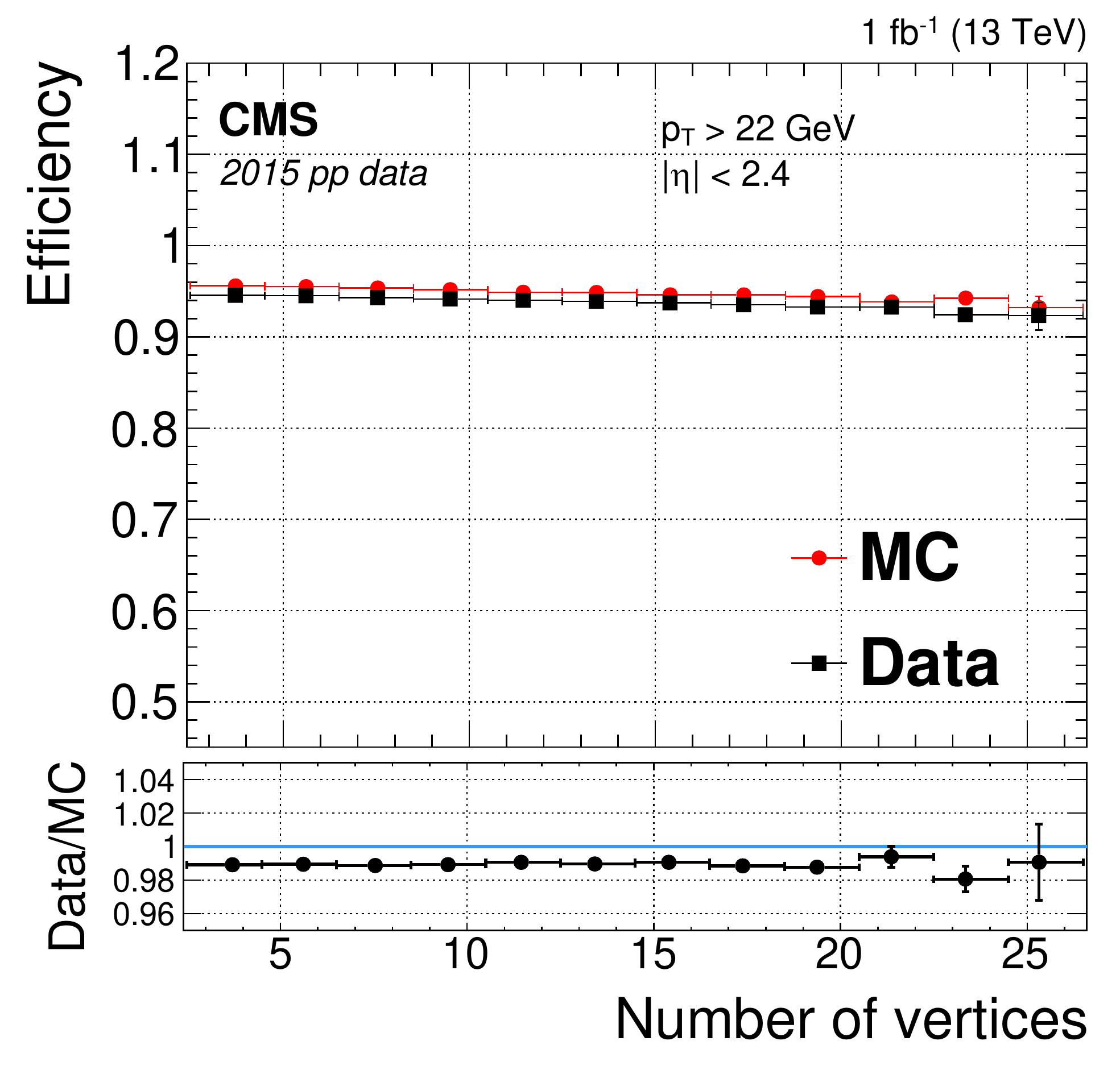}
  \caption{Isolated single-muon trigger efficiency measured with 2015 data (squares),
    simulation (circles), and the ratio (bottom inset).
    Results are plotted as a function of offline reconstructed muon \pt (upper left),
    $\eta$ (upper right), and number of primary vertices (lower).
    The statistical uncertainties in these values are smaller than the marker size in
    the figure.
  }
  \label{fig:HLTTrigEff}
\end{figure}

As described in Section~\ref{subsec:recoeff}, the efficiency of the trigger is measured
with the tag-and-probe technique.
In order not to bias the measurement of the trigger efficiency, the tag is geometrically
matched to the HLT trigger that selected the event.
In addition, it is also required to satisfy tight ID and PF isolation criteria in order
to reduce backgrounds.
The requirements on the probe are then tuned according to the reconstruction and
isolation criteria used in the analysis.
As an example, an analysis of muons with tight ID and PF isolation requirements might
use the isolated single-muon trigger to select events.
In this case, the probe muon is required to satisfy tight ID and PF isolation
requirements as per the analysis.
Using this technique, the efficiency of the isolated single-muon trigger with HLT \pt
threshold 20\GeV is shown in Fig.~\ref{fig:HLTTrigEff}.
The efficiency as a function of reconstructed muon \pt (Fig.~\ref{fig:HLTTrigEff}a)
rises sharply at the threshold.
Above 22\GeV, the inefficiency of a few percent is primarily caused by the L1 trigger
and the relative isolation criteria (see Tab.~\ref{tab:triggereff}).
Variations in efficiency as a function of $\eta$ (Fig.~\ref{fig:HLTTrigEff}b) are caused
by geometrical features of the detector that affect the L1 trigger efficiency.
The isolation requirement is responsible for the mild efficiency drop as a function of
the number of offline reconstructed vertices (Fig.~\ref{fig:HLTTrigEff}c).
The systematic uncertainty is estimated to be 0.5\% based on methods similar to those
described in Section~\ref{sec:efficiency}.
The simulation is in reasonable agreement with the data over the full momentum range and
angular acceptance.

A breakdown of the contributions to the isolated single-muon trigger efficiency from L1,
HLT track reconstruction, and online isolation is presented in
Table~\ref{tab:triggereff}, together with the scale factors between data and simulation.
Numbers are separated into barrel and endcap regions, and are integrated over offline
reconstructed $\pt > 22\GeV$ for each row, the momentum range commonly used for CMS
physics analyses with isolated single-muon triggers.
The first two rows show the efficiency of L1 candidates (\pt threshold 16\GeV) computed
with respect to tight muons passing PF isolation criteria.
The second two rows show the efficiency of HLT reconstruction (\pt threshold 20\GeV)
computed with respect to offline muons geometrically matched to L1 candidates, which
are used as the seeds for HLT tracking.
The use of two complementary reconstruction algorithms results in an efficiency exceeding
99\% for the HLT reconstruction of isolated single-muon triggers for prompt muons passing
L1 trigger requirements.
The last two rows show the effect of isolation on top of the HLT reconstruction.

\begin{table}
  \topcaption{Contributions to the isolated single-muon trigger efficiency in 2015
    data, integrated over $\pt > 22\GeV$.
    The first two rows show the level-1 efficiency (\pt threshold 16\GeV) with
    respect to offline muons.
    The second two rows show the HLT efficiency (\pt threshold 20\GeV) with
    respect to offline muons geometrically matched to L1 candidates.
    The last two rows show the online isolation efficiency with respect to offline
    muons firing HLT.
    The uncertainties in these values are statistical.
  }
  \centering
  \begin{tabular}{lccc}
    \hline
    Step
    & $\abs{\eta}$ region      &  Data eff. [\%]  &    Scale factor      \\
    \hline
    \multirow{2}{*}{L1 w.r.t. offline}
       & $0.0 < \abs{\eta} < 0.9$ & $96.86 \pm 0.02$ &  $0.9914 \pm 0.0005$ \\
    & $0.9 < \abs{\eta} < 2.4$ & $94.38 \pm 0.02$ &  $0.9947 \pm 0.0005$ \\[\cmsTabSkip]
    \multirow{2}{*}{HLT w.r.t L1}
       & $0.0 < \abs{\eta} < 0.9$ & $99.67 \pm 0.02$ &  $0.9967 \pm 0.0005$ \\
    & $0.9 < \abs{\eta} < 2.4$ & $99.46 \pm 0.02$ &  $0.9957 \pm 0.0005$ \\[\cmsTabSkip]
    \multirow{2}{*}{Online isolation w.r.t HLT}
       & $0.0 < \abs{\eta} < 0.9$ & $97.95 \pm 0.02$ &  $0.9906 \pm 0.0005$ \\
    & $0.9 < \abs{\eta} < 2.4$ & $98.28 \pm 0.02$ &  $0.9931 \pm 0.0005$ \\
    \hline
  \end{tabular}
  \label{tab:triggereff}
\end{table}

The efficiency for isolated single-muon triggers is improved with respect to
Run~1~\cite{POGPerformance,L1Paper} as a result of a combination of the changes in
HLT algorithms, the addition of RPC and CSC chambers in station~4, and the removal
of the ganging of strips in ME1/1a.
In the endcaps, the improvement in trigger efficiency (L1+HLT+isolation) relative to
the end of Run~1~\cite{L1Paper} is about 10\% for $\abs{\eta} > 1.2$ but reaches 20\% for
$\abs{\eta} \approx 2.4$.

A comparison of trigger rates at the same instantaneous luminosity and threshold
($\pt > 24\GeV$), and integrated over $\abs{\eta} < 2.4$, shows an increase of about 75\%
from Run~1 to Run~2.
This increase is approximately consistent with the increase of the inclusive
production cross sections for {\PW} and {\cPZ} bosons due to the change from
$\sqrt{s}=8\TeV$~\cite{PhysRevLett.112.191802} to
$\sqrt{s}=13\TeV$~\cite{CMS-PAS-SMP-15-004} with an additional
contribution from the increase in efficiency described above.

The combination of the updated HLT algorithms and the overall increase of HLT output
rate, together with a different allocation of the bandwidth, made it possible to
reduce the \pt thresholds on isolated single-muon triggers from 24\GeV in 2012 to
20\GeV in 2015.
The nonisolated triggers operated in 2015 with \pt thresholds of 45\GeV
($\abs{\eta} < 2.1$) and 50~\GeV ($\abs{\eta} < 2.4$).
For inclusive double-muon triggers, the use of track-based isolation requirements in
Run~2 resulted in a reduction of the rates of these triggers with respect to Run~1
despite the increase in collision energy.
In 2015, the thresholds for the two muons in double-muon triggers were
18\GeV and 7\GeV, the same values as in 2012.

\begin{figure}[ht]
  \centering
  \includegraphics[width=0.8\linewidth]{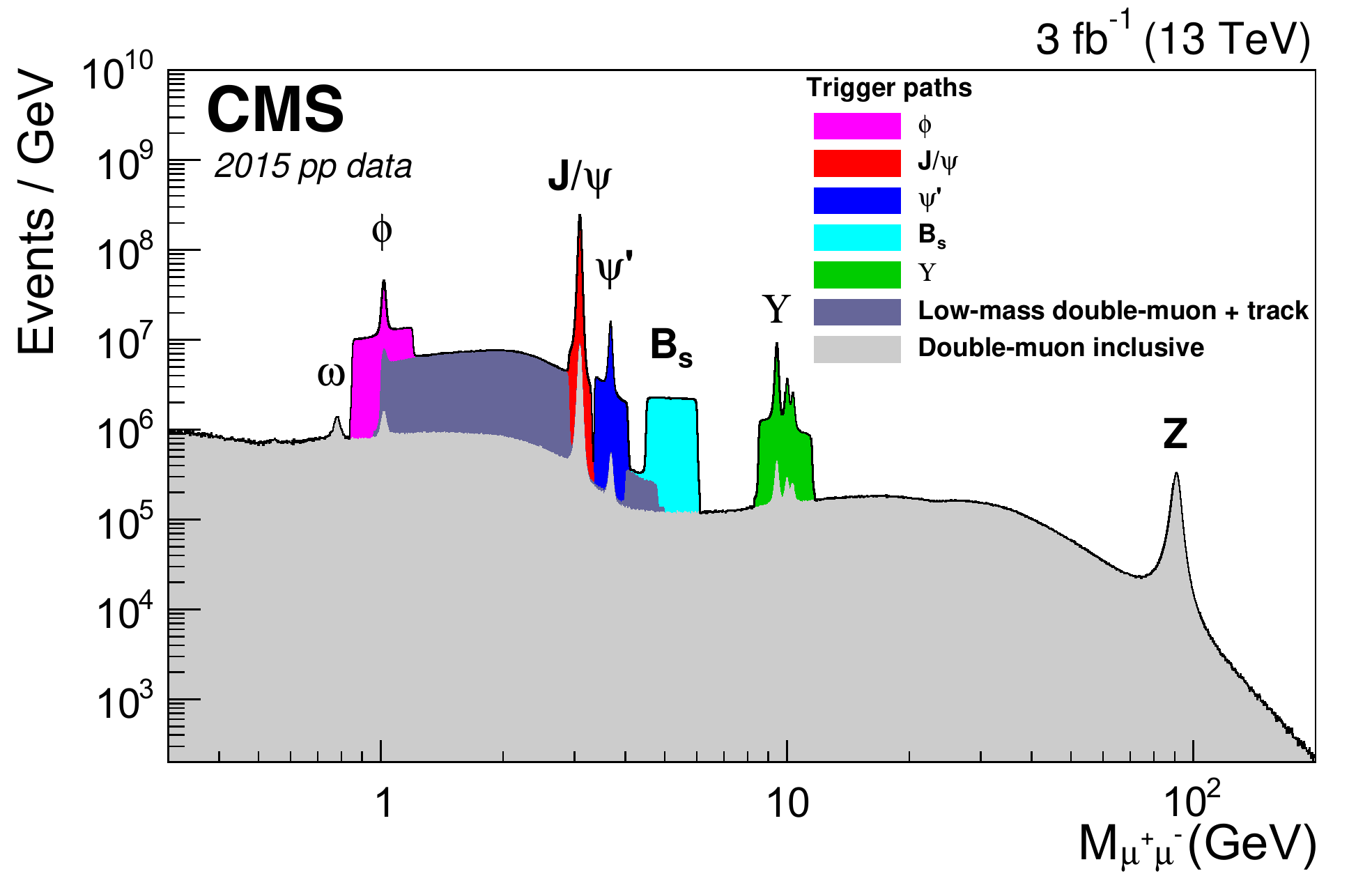}
  \caption{The dimuon invariant mass distribution reconstructed by the CMS HLT.
    Data were collected in 2015 with the inclusive double-muon trigger
    algorithm (gray), as well as triggers dedicated to selecting resonances at
    low masses.
  }
  \label{fig:HLTDiMuMass}
\end{figure}

Figure~\ref{fig:HLTDiMuMass} shows the invariant mass distribution of oppositely
charged muon pairs selected by the inclusive trigger on isolated double-muons.
The x-axis is logarithmic so the entries are scaled to the width of each bin.
Data are also included from specific double-muon triggers tuned to select resonances
at low invariant mass.
The figure clearly demonstrates the ability of CMS to identify muons, trigger on them,
and reconstruct the muon kinematics to unambiguously identify particles that decay into
muons over a broad energy range.

\section{Summary}
\label{sec:summary}

The performance of the CMS muon detector and reconstruction software has been studied
using data from proton-proton collisions at center-of-mass energy $\sqrt{s}=13\TeV$,
collected in 2015 and 2016 during LHC Run~2.
These results are compared to the previously published results collected in 2010 at
$\sqrt{s}=7\TeV$ with instantaneous luminosities about a factor of 40 lower.
Important modifications to many components of the muon system were made before Run~2
in anticipation of the higher collision energy and the increased luminosity.
These included modifications to drift tubes, cathode strip chambers, and resistive
plate chambers, as well as improved algorithms for the high-level trigger and offline
reconstruction.
Although not comprehensive, a set of representative figures of merit for the system
performance include:
\begin{itemize}
\item reconstructed hit spatial resolution $\approx 50-300$\mum;
\item reconstructed hit efficiency  $\approx 94-99$\%;
\item segment timing resolution $< 3$\unit{ns};
\item segment efficiency  $\approx 97$\%;
\item trigger bunch crossing identification $> 99$\%;
\item trigger efficiency $> 90$\%;
\item muon timing resolution  $\approx 1.4$\unit{ns};
\item muon reconstruction and identification efficiency $> 96$\%;
\item muon isolation efficiency $> 95$\%.
\end{itemize}

As a result of the improvements to the detector and the reconstruction algorithms,
and despite the higher luminosity and pileup in Run~2, the muon performance is
better than, or at least as good as, it was in 2010.
Detector performance remains within the design specifications and the muon
reconstruction results are well reproduced by Monte Carlo simulation.

\begin{acknowledgments}
\hyphenation{Bundes-ministerium Forschungs-gemeinschaft Forschungs-zentren Rachada-pisek} We congratulate our colleagues in the CERN accelerator departments for the excellent performance of the LHC and thank the technical and administrative staffs at CERN and at other CMS institutes for their contributions to the success of the CMS effort. In addition, we gratefully acknowledge the computing centers and personnel of the Worldwide LHC Computing Grid for delivering so effectively the computing infrastructure essential to our analyses. Finally, we acknowledge the enduring support for the construction and operation of the LHC and the CMS detector provided by the following funding agencies: the Austrian Federal Ministry of Science, Research and Economy and the Austrian Science Fund; the Belgian Fonds de la Recherche Scientifique, and Fonds voor Wetenschappelijk Onderzoek; the Brazilian Funding Agencies (CNPq, CAPES, FAPERJ, and FAPESP); the Bulgarian Ministry of Education and Science; CERN; the Chinese Academy of Sciences, Ministry of Science and Technology, and National Natural Science Foundation of China; the Colombian Funding Agency (COLCIENCIAS); the Croatian Ministry of Science, Education and Sport, and the Croatian Science Foundation; the Research Promotion Foundation, Cyprus; the Secretariat for Higher Education, Science, Technology and Innovation, Ecuador; the Ministry of Education and Research, Estonian Research Council via IUT23-4 and IUT23-6 and European Regional Development Fund, Estonia; the Academy of Finland, Finnish Ministry of Education and Culture, and Helsinki Institute of Physics; the Institut National de Physique Nucl\'eaire et de Physique des Particules~/~CNRS, and Commissariat \`a l'\'Energie Atomique et aux \'Energies Alternatives~/~CEA, France; the Bundesministerium f\"ur Bildung und Forschung, Deutsche Forschungsgemeinschaft, and Helmholtz-Gemeinschaft Deutscher Forschungszentren, Germany; the General Secretariat for Research and Technology, Greece; the National Research, Development and Innovation Fund, Hungary; the Department of Atomic Energy and the Department of Science and Technology, India; the Institute for Studies in Theoretical Physics and Mathematics, Iran; the Science Foundation, Ireland; the Istituto Nazionale di Fisica Nucleare, Italy; the Ministry of Science, ICT and Future Planning, and National Research Foundation (NRF), Republic of Korea; the Lithuanian Academy of Sciences; the Ministry of Education, and University of Malaya (Malaysia); the Mexican Funding Agencies (BUAP, CINVESTAV, CONACYT, LNS, SEP, and UASLP-FAI); the Ministry of Business, Innovation and Employment, New Zealand; the Pakistan Atomic Energy Commission; the Ministry of Science and Higher Education and the National Science Centre, Poland; the Funda\c{c}\~ao para a Ci\^encia e a Tecnologia, Portugal; JINR, Dubna; the Ministry of Education and Science of the Russian Federation, the Federal Agency of Atomic Energy of the Russian Federation, Russian Academy of Sciences and the Russian Foundation for Basic Research; the Ministry of Education, Science and Technological Development of Serbia; the Secretar\'{\i}a de Estado de Investigaci\'on, Desarrollo e Innovaci\'on, Programa Consolider-Ingenio 2010, Plan Estatal de Investigaci\'on Cient\'{\i}fica y T\'ecnica y de Innovaci\'on 2013-2016, Plan de Ciencia, Tecnolog\'{i}a e Innovaci\'on 2013-2017 del Principado de Asturias and Fondo Europeo de Desarrollo Regional, Spain; the Swiss Funding Agencies (ETH Board, ETH Zurich, PSI, SNF, UniZH, Canton Zurich, and SER); the Ministry of Science and Technology, Taipei; the Thailand Center of Excellence in Physics, the Institute for the Promotion of Teaching Science and Technology of Thailand, Special Task Force for Activating Research and the National Science and Technology Development Agency of Thailand; the Scientific and Technical Research Council of Turkey, and Turkish Atomic Energy Authority; the National Academy of Sciences of Ukraine, and State Fund for Fundamental Researches, Ukraine; the Science and Technology Facilities Council, UK; the US Department of Energy, and the US National Science Foundation.

Individuals have received support from the Marie-Curie program and the European Research Council and Horizon 2020 Grant, contract No. 675440 (European Union); the Leventis Foundation; the A. P. Sloan Foundation; the Alexander von Humboldt Foundation; the Belgian Federal Science Policy Office; the Fonds pour la Formation \`a la Recherche dans l'Industrie et dans l'Agriculture (FRIA-Belgium); the Agentschap voor Innovatie door Wetenschap en Technologie (IWT-Belgium); the F.R.S.-FNRS and FWO (Belgium) under the ``Excellence of Science - EOS" - be.h project n. 30820817; the Ministry of Education, Youth and Sports (MEYS) of the Czech Republic; the Lend\"ulet (``Momentum") Programme and the J\'anos Bolyai Research Scholarship of the Hungarian Academy of Sciences, the New National Excellence Program \'UNKP, the NKFIA research grants 123842, 123959, 124845, 124850 and 125105 (Hungary); the Council of Scientific and Industrial Research, India; the HOMING PLUS program of the Foundation for Polish Science, cofinanced from European Union, Regional Development Fund, the Mobility Plus program of the Ministry of Science and Higher Education, the National Science Center (Poland), contracts Harmonia 2014/14/M/ST2/00428, Opus 2014/13/B/ST2/02543, 2014/15/B/ST2/03998, and 2015/19/B/ST2/02861, Sonata-bis 2012/07/E/ST2/01406; the National Priorities Research Program by Qatar National Research Fund; the Programa de Excelencia Mar\'{i}a de Maeztu and the Programa Severo Ochoa del Principado de Asturias; the Thalis and Aristeia programs cofinanced by EU-ESF and the Greek NSRF; the Rachadapisek Sompot Fund for Postdoctoral Fellowship, Chulalongkorn University and the Chulalongkorn Academic into Its 2nd Century Project Advancement Project (Thailand); the Welch Foundation, contract C-1845; and the Weston Havens Foundation (USA).
\end{acknowledgments}

\bibliography{auto_generated}

\appendix

\cleardoublepage \appendix\section{The CMS Collaboration \label{app:collab}}\begin{sloppypar}\hyphenpenalty=5000\widowpenalty=500\clubpenalty=5000\vskip\cmsinstskip
\textbf{Yerevan Physics Institute,  Yerevan,  Armenia}\\*[0pt]
A.M.~Sirunyan,  A.~Tumasyan
\vskip\cmsinstskip
\textbf{Institut f\"{u}r Hochenergiephysik,  Wien,  Austria}\\*[0pt]
W.~Adam,  F.~Ambrogi,  E.~Asilar,  T.~Bergauer,  J.~Brandstetter,  E.~Brondolin,  M.~Dragicevic,  J.~Er\"{o},  A.~Escalante Del Valle,  M.~Flechl,  M.~Friedl,  R.~Fr\"{u}hwirth\cmsAuthorMark{1},  V.M.~Ghete,  J.~Grossmann,  J.~Hrubec,  M.~Jeitler\cmsAuthorMark{1},  A.~K\"{o}nig,  N.~Krammer,  I.~Kr\"{a}tschmer,  D.~Liko,  T.~Madlener,  I.~Mikulec,  E.~Pree,  N.~Rad,  H.~Rohringer,  J.~Schieck\cmsAuthorMark{1},  R.~Sch\"{o}fbeck,  M.~Spanring,  D.~Spitzbart,  A.~Taurok,  W.~Waltenberger,  J.~Wittmann,  C.-E.~Wulz\cmsAuthorMark{1},  M.~Zarucki
% ...
\vskip\cmsinstskip
% ...
1:  Also at Vienna University of Technology,  Vienna,  Austria\\
% ...
\end{sloppypar}
\end{document}